\documentclass[
superscriptaddress,
reprint,
amsmath,
floatfix,
pra
]{revtex4-1}

\usepackage{mathtools,amssymb,lipsum}
\usepackage{booktabs}
\usepackage{siunitx}
\usepackage[]{hyperref}
\usepackage{multirow}
\usepackage{bbold} 
\usepackage[normalem]{ulem}
\usepackage{csvsimple}
\usepackage{physics}
\usepackage{xcolor}
\usepackage{printlen}
\usepackage{tabularx}
\usepackage{tikz}
\usepackage{subfiles}

\usetikzlibrary{arrows}
\usetikzlibrary{decorations.pathmorphing}

\tikzset{snake it/.style={decorate, decoration=snake}}

\newcommand{\thetitle}{Quantum theory for phonon lasing and non-classical state generation in mixed-species and single trapped ions}
\newcommand{\gh}{g_\mathrm{h}}
\newcommand{\gc}{g_\mathrm{c}}
\newcommand{\gamh}{\gamma_\mathrm{h}}
\newcommand{\gamc}{\gamma_\mathrm{c}}

\newcommand{\sph}{\sigma^\mathrm{h}_\mathrm{+}}
\newcommand{\smh}{\sigma^\mathrm{h}_\mathrm{-}}
\newcommand{\spc}{\sigma^\mathrm{c}_\mathrm{+}}
\newcommand{\smc}{\sigma^\mathrm{c}_\mathrm{-}}
\newcommand{\szh}{\sigma^\mathrm{h}_\mathrm{z}}
\newcommand{\szc}{\sigma^\mathrm{c}_\mathrm{z}}
\newcommand{\nuh}{\eta_\mathrm{h}}
\newcommand{\nuc}{\eta_\mathrm{c}}
\newcommand{\kaph}{\kappa_\mathrm{h}}
\newcommand{\kapc}{\kappa_\mathrm{c}}

\newcommand{\rhn}[2]{\rho_\mathrm{#1}^\mathrm{#2}}
\newcommand{\rhnd}[2]{\dot{\rho}_\mathrm{#1}^\mathrm{#2}}
\newcommand{\gzn}{g^\mathrm{(2)}(0)}
\newcommand{\sto}{\sigma_\mathrm{21}}
\newcommand{\sot}{\sigma_\mathrm{12}}

\newcommand{\theauthors}{
\setcounter{affil}{0}
\author{D. Baur}
\email{davbaur@phys.ethz.ch}
\affiliation{Institute for Quantum Electronics, ETH Z\"urich, Otto-Stern-Weg 1, 8093 Z\"urich, Switzerland}
\affiliation{Quantum Center, ETH Z{\"u}rich, 8093 Z{\"u}rich, Switzerland}
\author{T. Behrle} 
\thanks{Present address: ETH Z\"urich – PSI Quantum Computing Hub, Paul Scherrer Institut (PSI), 5232 Villigen, Switzerland}
\affiliation{Institute for Quantum Electronics, ETH Z\"urich, Otto-Stern-Weg 1, 8093 Z\"urich, Switzerland}
\affiliation{Quantum Center, ETH Z{\"u}rich, 8093 Z{\"u}rich, Switzerland}
\author{I. Rojkov}
\thanks{Present address: Department of Physics, Yale University, New Haven, Connecticut 06520, USA}
\affiliation{Institute for Quantum Electronics, ETH Z\"urich, Otto-Stern-Weg 1, 8093 Z\"urich, Switzerland}
\affiliation{Quantum Center, ETH Z{\"u}rich, 8093 Z{\"u}rich, Switzerland}
\author{J. Jeske}
\affiliation{Fraunhofer Institute for Applied Solid State Physics IAF, Tullastra{\ss}e 72, 79108 Freiburg, Germany}
\author{S. Yelin}
\affiliation{Harvard University, 17 Oxford Street, Cambridge, MA 02138, USA}
\author{J. Home}
\affiliation{Institute for Quantum Electronics, ETH Z\"urich, Otto-Stern-Weg 1, 8093 Z\"urich, Switzerland}
\affiliation{Quantum Center, ETH Z{\"u}rich, 8093 Z{\"u}rich, Switzerland}
\author{F. Reiter}
\affiliation{Institute for Quantum Electronics, ETH Z\"urich, Otto-Stern-Weg 1, 8093 Z\"urich, Switzerland}
\affiliation{Quantum Center, ETH Z{\"u}rich, 8093 Z{\"u}rich, Switzerland}
\affiliation{Fraunhofer Institute for Applied Solid State Physics IAF, Tullastra{\ss}e 72, 79108 Freiburg, Germany}
}

\begin{document}

\title{\thetitle}
\theauthors

\begin{abstract}
In this article we present a comprehensive theoretical investigation of phonon lasing with mixed-species trapped ions, as demonstrated in [T. Behrle, Phys. Rev. Lett. 131 (2023)], employing both a semi-classical mean-field description and a full quantum theory. We derive an analytic expression for the second-order coherence function, confirming the experimental observation of the system's lasing behaviour above threshold. Building on the successful implementation of the two-ion lasing scheme, we propose a novel approach for achieving phonon lasing with a single trapped ion, offering significant experimental advantages and making the implementation of multiple phonon lasers within a single setup feasible. Furthermore, we explore lasing in a squeezed basis and in different regimes of the Lamb-Dicke approximation, highlighting the potential to produce non-classical states with promising applications in precision sensing. Our analysis of a sensing protocol based on squeezed states, using experimentally feasible parameters, shows a sensitivity enhancement of up to two orders of magnitude.
\end{abstract}

\maketitle
\section{Introduction}
Lasers are archetypal examples of open spin-boson systems. Analogous lasing dynamics can be observed when replacing the electromagnetic mode with that of a mechanical oscillator, yielding a phonon laser. These systems serve as versatile platforms for investigating fundamental physics, such as synchronization phenomena~\cite{Sheng2020Self,Hush2015Spin} or as broadly tunable quantum simulators~\cite{Behrle2023}. Additional proposed applications include the realization of $\mathcal{PT}$-symmetric and non-reciprocal phonon lasers, which facilitate the exploration of $\mathcal{PT}$-symmetric concepts, topological and chiral acoustics \cite{Jiang2018Nonreciprocal,Jing2014PT}.
Beyond fundamental demonstrations, lasers have been widely employed as precision sensing devices~\cite{Jeske_2016, RamanNair_2021, Dumeige19, Webb2021LaserThreshold}. Similarly, phonon lasers can offer analogous high-precision sensing capabilities \cite{Porras2017, Liu2021Phonon, Wei2022}. 
Experimental realizations of phonon lasers span a diverse range of platforms, including trapped ions~\cite{Vahala2009, inj_locking, Ip2018Phonon, Lee2023Prototype, Xie2013Pulsed, Behrle2023}, nanomechanical resonators~\cite{Zhang2021, Kemiktarak2014, Cohen2015}, optical tweezers~\cite{Pettit2019}, microcavities~\cite{Grudinin2010}, and electromechanical systems~\cite{Wen2020, Mahboob2013, Beardsley2010}.
\begin{figure}[!h]
    \centering
    \includegraphics[width=0.45\textwidth]{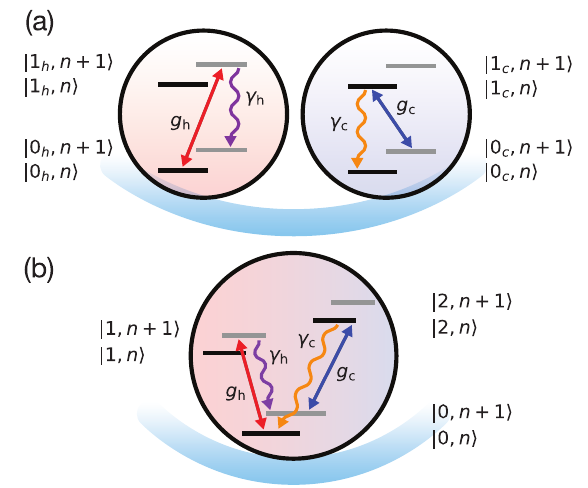}
    \caption{(a): Two-ion phonon lasing scheme. On the first species (index h) a Anti-Jaynes Cummings (AJC) interaction is shown in red with effective driving strength $\gh$. It couples the ion spin to the common motional mode, i.e. it creates an interaction between the states $\ket{0_\mathrm{h},n}$ and $\ket{1_\mathrm{h},n+1}$. The engineered ion decay is given in purple with a rate $\gamh$. The combined effective action on this ion is the addition of phonons, i.e. heating, hence the index ``h'' and red color scheme. On the second species (index c) a Jaynes-Cummings (JC) interaction is given in blue with strength $\gc$. It couples the states $\ket{1_\mathrm{c},n}$ and $\ket{0_\mathrm{c},n+1}$. The engineered ion decay is given in yellow with rate $\gamc$. The index ``c'' and blue color scheme reflect the combined effective action of this species on the motional mode:  cooling. (b): The single-ion lasing scheme uses three internal levels to realise similar lasing dynamics. Between the ground state and one of the excited states a blue (red) sideband is implemented as well as engineered decays from these excited states.}
    \label{fig:twoIonScheme}
\end{figure}

Additionally, implementations involving quantum dots coupled to nanowires and THz acoustic nanocavities have been proposed \cite{Khaetskii2013, Kabuss2012}. Recently, a novel type of quantum simulator was implemented and studied in a mixed-species trapped ion setup, operating as a phonon laser in the few-quanta regime \cite{Behrle2023}. While a well-established quantum theory exists for photon lasing \cite{Scully1967}, developing a quantum theory for this phonon laser has proven more challenging, particularly in describing key aspects, such as the second-order coherence function \cite{Glauber1963}. This is necessary to confirm the coherent nature of the lasing state.

This article aims to address this shortcoming and extends the existing mean-field treatment of the phonon lasing scheme to a full quantum theory. We propose extensions of the experimentally studied model, such as squeezed lasing dynamics and the utilization of higher-order terms in the Lamb-Dicke expansion as a saturation mechanism and as a means to reach non-classical steady states.

The realization of a phonon laser using mixed-species trapped ion chains is experimentally challenging, as it involves operating two separate laser systems and the simultaneous control of two ion species. This complicates the extension to a system of multiple coupled lasers. To address this challenge, we investigate a novel lasing scheme based on a single trapped ion, capable of analogous lasing dynamics as the two-ion model (see Fig.~\ref{fig:twoIonScheme} for an overview of both schemes). We study this new model with the same theories as the two-ion model and discuss possible extensions. Specifically, we investigate the usage of the lasing threshold for enhanced sensing of mechanical oscillations. This technique is already routinely used for photon based lasers~\cite{Jeske_2016, RamanNair_2021, Dumeige19, Webb2021LaserThreshold}. Additionally we explore how phonon lasing in a squeezed mode as well as the non-classical states generated by the use of the higher-order Lamb-Dicke terms could open up new ways of realising quantum advantage in sensing. We quantitatively explore the sensing enhancement when employing a squeezing scheme with experimentally achievable parameters.

This article is organized as follows: in Section~\ref{sec:twoionPLM} we present the two-ion phonon lasing scheme and investigate it in the mean-field approximation, as a first step to understanding the different phases the system can attain. We then move to a full quantum description of the system, which allows us to investigate the phonon statistics of the steady state. We discuss the possibility of using higher-order terms of the Lamb-Dicke expansion as a saturation mechanism and as a means of engineering non-classical states. Lastly, we extend the model to include squeezed lasing dynamics and discuss the applications of the squeezed phonon laser in sensing. In Section~\ref{sec:singelIonPLM}, we propose a novel lasing scheme, which can be implemented on a single trapped ion. We show through a mean-field analysis, a quantum theory, and numerical simulations that in this simplified setup analogous lasing dynamics to those of the two-ion model can be observed. We investigate the effect of violating the first-order Lamb-Dicke approximation and its application as a saturation mechanism and show the implementation of squeezed lasing. In Section~\ref{sec:Comparison}, we compare the single and two-ion lasing models, and discuss the physical origin of differences between them.

\section{Two-ion phonon lasing model}
\label{sec:twoionPLM}
The two-ion phonon lasing model, illustrated in Fig.~\ref{fig:twoIonScheme}~(a), comprises two effective two-level systems, referred to as spins, defined by the electronic states of the single valence electrons of each ion. Additionally, the model includes a shared harmonic mode, corresponding to one of the six motional modes the two ions share in a harmonic trapping potential. The spin states are described by Pauli operators, denoted by $\sigma_\mathrm{x,y,z}$ and raising (lowering) operators $\sigma_\mathrm{\pm}=\sigma_\mathrm{x}\pm i \sigma_\mathrm{y}$. The harmonic motional mode is described by the bosonic creation (destruction) operator $a^\dagger$ ($a$). The interaction between the spins and the motion of the ions is modeled with a Jaynes-Cummings (JC) Hamiltonian and an anti-Jaynes Cummings (AJC) Hamiltonian
\begin{align}
    H_\mathrm{h}&=\gh (a^\dagger \sph + a \smh),\label{twoion:Hh}\\
    H_\mathrm{c}&=\gc (a^\dagger \smc + a \spc),\label{twoion:Hc}
\end{align}
where $\gh$ is the driving strength given as $\nuh\Omega_\mathrm{h}\exp(-\nuh^2/2)$, with $\nuh$ the Lamb-Dicke parameter and $\Omega_\mathrm{h}$ the Rabi frequency (analogous for the other species with index $c$). 

Additionally, a decay channel is implemented in each of the ions with time constants $\gamh$ and $\gamc$, respectively. We introduce these channels into the description with the Lindblad jump operators
\begin{align}
    L_\mathrm{h}=\sqrt{\gamh}\smh \quad \text{and}\quad L_\mathrm{c}=\sqrt{\gamc}\smc.\label{twoion:Lh}
\end{align}
The ion species which evolves under the JC interaction and the engineered decay effectively lowers the occupation of the motional mode. It is therefore referred to as the cooling ion (index c). The ion species which evolves under the AJC interaction and the decay channel effectively increases the occupation of the motional mode and is referred to as the heating ion (index h). 

With these two dissipation channels and the two coherent terms given in Eqs.~\eqref{twoion:Hh} and \eqref{twoion:Hc} we can describe the lasing dynamics which can be observed in a two-ion lasing scheme. Note that by introducing incoherent channels from $\ket{0}$ to $\ket{1}$ on both ions, the system can be mapped onto standard lasing in an inverted two-level basis, where the role of heating and cooling is interchanged between the two ions.

\subsection{Mean-field analysis}
\label{subsec:twoion:mf}
In this first subsection, we investigate the two-ion model using a mean-field approach in a semi-classical limit, giving access to the steady-state mean phonon number. Based on these results, we identify the different phases of the phonon laser. For this derivation, we assume that the steady-state phonon occupation is at a high average value $\expval{n}\gg1$ and that the variance of this state is large compared to the Fock level spacing $\expval{\Delta n}\gg1$ \cite{scully_zubairy_1997}. We can therefore assume that the time evolution of one JC or AJC cycle on either ion does not considerably change the average phonon value and its variance. This allows us to decouple the dynamics of the spin from the motion of the ion. We further make use of the first-order cumulant assumption for all system variables; for example
\begin{align}
    \expval{\sigma_\mathrm{n}\sigma_\mathrm{m}}=\expval{\sigma_\mathrm{n}}\expval{\sigma_\mathrm{m}} \hspace{.5cm} n,m\in \{x,y,z,+,-\},
\end{align}
which for the phonon field is equivalent to assuming a coherent state.
We set up the dynamical equation for the motional mode and the spin variables using the von Neumann equation
\begin{align}
    \frac{d}{dt} \expval{O} = i \expval{[H,O]} + \sum_k \expval{\tilde{\mathcal{D}}[L_k](O)},\label{eq:2ion:mf:oeevolution}
\end{align}
with $H=H_\mathrm{h}+H_\mathrm{c}$ and $L_\mathrm{k}\in\{L_\mathrm{h},L_\mathrm{c}\}$. This results in the following set of coupled equations
\begin{align}
    \frac{d}{dt}\expval{a} =& -i\gc\expval{\smc} -i\gh\expval{\sph},\label{eq:2ion:mf:adot}\\
    \frac{d}{dt}\expval{\sph} =& -\frac{\gamh}{2}\expval{\sph}-i\gh\expval{a}\expval{\szh},\\
    \frac{d}{dt}\expval{\spc} =& -\frac{\gamc}{2}\expval{\spc}-i\gc\expval{a^\dagger}\expval{\szc},\\
    \frac{d}{dt}\expval{\szh} =& 
 2i\gh\left(\expval{a}\expval{\smh} - \expval{a^\dagger}\expval{\sph}\right)\nonumber\\
    & -\gamh\left(\expval{\szh}+1\right),\\
    \frac{d}{dt}\expval{\szc} =& 
 2i\gc\left(\expval{a^\dagger}\expval{\smc} - \expval{a}\expval{\spc}\right)\nonumber\\
    & -\gamc\left(\expval{\szc}+1\right).
\end{align}
We solve the dynamical equations for the spins under the assumption that the expectation values of the motional mode $\expval{a}$ and $\expval{a^\dagger}$ are constant, based on the separation of the time scale between spin and motional evolution. From this, we find the following expressions for the spin expectation values in the steady state
\begin{align}
    \expval{\sph}&=\frac{2i\frac{\gh}{\gamh} \expval{a}}{1+8\frac{\gh^2}{\gamh^2}I}\quad \text{and} \quad
    \expval{\spc}=\frac{2i\frac{\gc}{\gamc} \expval{a^\dagger}}{1+8\frac{\gc^2}{\gamc^2}I},\label{eq:spin:eqvalue}
\end{align}
where $I=\expval{a^\dagger}\expval{a}$, the phonon laser intensity or equivalently the average phonon number. Note that this derivation implicitly made the assumption that the phonon mode is in a coherent state. We will discuss in the next subsection a different approach to verify that this condition is indeed fulfilled. Inserting Eqs.~\eqref{eq:spin:eqvalue} into Eq.~\eqref{eq:2ion:mf:adot} we find the evolution equation for the motional mode of the ions 
\begin{align}
    \frac{d}{dt}\expval{a}=\expval{a}\left(\frac{\frac{2\gh^2}{\gamh}}{1+8\frac{\gh^2}{\gamh^2} I}-\frac{\frac{2\gc^2}{\gamc}}{1+8\frac{\gc^2}{\gamc^2}I}\right).
\end{align}
In a similar way, we derive the dynamical equation for $\expval{a^\dagger}$. With these two equations and the identity $\dot{I}=\expval{\dot{a}}\expval{a^\dagger}+\expval{a}\expval{\dot{a}^\dagger}$, we find the following evolution equation for the phonon laser intensity
\begin{align}
    \frac{d}{dt}I=2I\left(\frac{\frac{2\gh^2}{\gamh}}{1+8\frac{\gh^2}{\gamh^2} I}-\frac{\frac{2\gc^2}{\gamc}}{1+8\frac{\gc^2}{\gamc^2}I}\right).\label{twoion:master:I}
\end{align}
This equation allows us to identify the different phases and corresponding steady states. The first steady state is given by $I=0$. As no phonons are excited, we refer to this solution as the ``dark'' phase, which is stable if $\gamc>\gamh$ and $\kaph:=\frac{\gh^2}{\gamh}<\frac{\gc^2}{\gamc}=:\kapc$, i.e. if the effective cooling rate exceeds the effective heating rate and any population at a non-zero $n$ is brought back to the vacuum state. This state becomes unstable if $\kaph>\kapc$ and the system transitions into a steady-state lasing phase, characterized by a non-zero $I$. This transition corresponds to the well-known lasing phase transition at $C=1$ \cite{scully_zubairy_1997}, with the cooperativity parameter in this case corresponding to $C=\frac{\kaph}{\kapc}$. Consequently, we refer to the condition $\kaph=\kapc$ as the lasing threshold. This new steady state is defined by the brackets in Eq.~\eqref{twoion:master:I} being equal to zero. This also corresponds to the intersection of the effective heating and cooling rate. We define these rates as
\begin{align}
    R_\mathrm{h}=\frac{2\kaph}{1+8\frac{\gh^2}{\gamh^2}I},
\end{align}
and analogously for the effective cooling rate $R_\mathrm{c}$. We translate this into a formula for the expected phonon number in the lasing steady state
\begin{align}
    I_\mathrm{ss}=\frac{\gamh^2\gamc^2\left(\kaph-\kapc\right)}{8\gh^2\gc^2(\gamc-\gamh)}.\label{eq:2ion:mf:n}
\end{align}
This state is stable if $\kaph>\kapc$ and $\gamc>\gamh$. If we violate the latter condition Eq.~\eqref{eq:2ion:mf:n} predicts a negative phonon number. We will later identify this unphysical result as a runaway phase, where the phonon number diverges.

\begin{figure}[ht]
    \centering
    \includegraphics[width=0.4\textwidth]{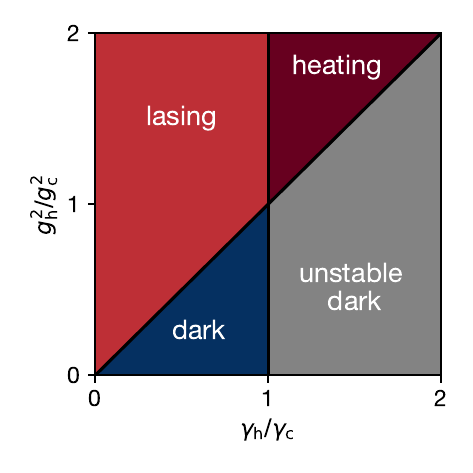}
    \caption{Phase diagram as obtained from mean-field equations. The y-axis is given by the ratio of the coherent coupling strengths of the JC and AJC Hamiltonian. The x-axis is given by the ratio of the incoherent decay rates of the two ions. In the lower left area, termed ``dark'', the laser is below the lasing threshold and stable at zero phonons. Crossing the diagonal line, given by $\kapc=\kaph$, we enter the stable lasing phase where a finite steady state is obtained. To the right of the vertical line, $\gamh=\gamc$, no stable solution is found: above the threshold, the phonon number diverges, resulting in a ``heating'' phase, below a zero steady state exists, which is not stable against perturbations. We term this state ``unstable dark''.}
    \label{fig:Two-ion:PD}
\end{figure}

With these results, we have found a broad overview of the parameter space of the system, as depicted in Fig.~\ref{fig:Two-ion:PD}. The axes are given by the ratio of the two coherent parameters ($\gh$,$\gc$) and the ratio of the two dissipation time scales ($\gamh$,$\gamc$), respectively. As discussed before, we find finite stable solutions for $n$, if $\gamh<\gamc$, which corresponds to the vertical line in Fig.~\ref{fig:Two-ion:PD}. The lasing threshold, given by $\kaph=\kapc$ is given by the diagonal line. Depending on the ratio $\gamh/\gamc$, we find below the lasing threshold, the dark and unstable dark phases, whereas above it we observe the stable lasing and heating phase.

\begin{figure}[ht]
    \centering
    \includegraphics[width=0.5\textwidth]{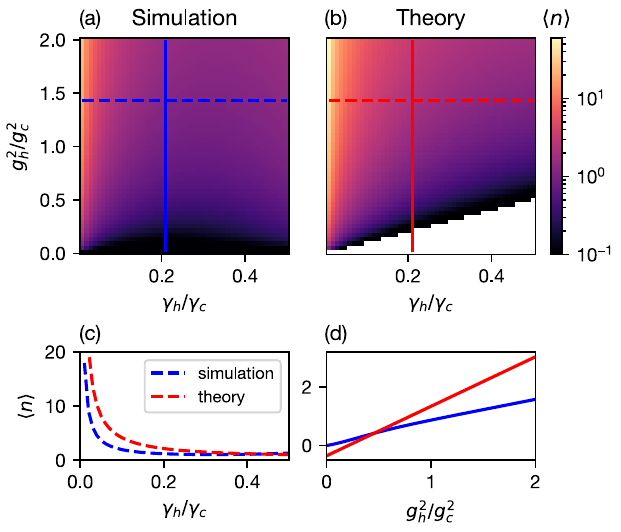}
    \caption{Comparison between the mean phonon number as simulated (a) and according to the mean-field theory (b); both in logarithmic scale. In the white shaded area (subplot (b)) the mean phonon number is negative according to Eq.~\eqref{eq:2ion:mf:n}. Here, the formula predicts a non-zero steady state, which is however unstable (as shown later). The stable state is the dark state, with zero phonon occupation in this parameter range, according to the mean-field analysis. In the simulation, we see a non-zero phonon occupation in this area. (c) and (d): Two cuts of the phase space are shown along the $\gamh/\gamc$ and $\gh^2/\gc^2$ direction, as indicated by the respectively colored dashed and solid lines in (a) and (b). For these plots the values of $\gh$ and $\gamh$ were fixed to $1.0$ and $1.5$ respectively and $\gc$ and $\gamc$ are scanned.}
    \label{fig:nComparison}
\end{figure}

We further compare the mean-field result of Eq.~\eqref{eq:2ion:mf:n} to the simulation of the system in Fig.~\ref{fig:nComparison}. The simulation numerically determines the steady states of the Liouvillian, described by Eqs.~\eqref{twoion:Hh}-\eqref{twoion:Lh}, for a given set of parameters.
The simulated results shown in Fig.~\ref{fig:nComparison} are found in agreement with the mean-field results discussed above. Notably, this agreement is present even in the regime where the steady state of the laser is occupied by only a few phonons, which would violate the assumption made in the derivation of this result ($\expval{n}\gg1$). According to simulation, the sharp lasing phase transition predicted by the mean-field theory, is in fact a smooth crossover from a dark state with zero phonons to a non-zero intensity above threshold. The negative values of the mean-field theory are predicted in a parameter range where the non-zero state is unstable (as discussed later), and the zero phonon occupation state is stable. We therefore mask the negative values (white area in Fig.~\ref{fig:nComparison}(b)).
This treatment gives us an overview of the phase diagram and an intuition for the microscopic processes leading to each of the four phases. The agreement between the simulation and the mean-field theory suggests that the phononic sub-system is indeed in a coherent state.

\subsection{Quantum theory}
In order to confirm the lasing nature of the steady state found in the mean-field analysis, we turn our attention to a quantum theory, which does not rely on the first-order mean-field assumption. We present an approach, which finds the phonon distribution by deriving a recurrence equation for $p(n)$, the phonon occupation probability at the Fock level $n$. Based on this we calculate the second-order coherence function $\gzn$ and confirm the lasing nature of the steady state above the lasing threshold.  

\subsubsection{Phonon distribution and coherence}\label{subsec:2ion:qt:pd}
To find an expression for the phonon number statistics of the state, we calculate the expectation value of the dynamical equations in terms of the Fock levels $\bra{n}$ and $\ket{n'}$. In this way, the information of the operator order is retained when manipulating the equations. For this derivation, we adopt the following notation based on Ref.~\cite{Scully1967} $\rho_\mathrm{xy,kl;nn'}:=\bra{x}_\mathrm{h}\bra{k}_\mathrm{c}\bra{n}_\mathrm{m}\rho\ket{n'}_\mathrm{m}\ket{l}_\mathrm{c}\ket{y}_\mathrm{h}$, $\rhn{xy;nn'}{h}:=\bra{x}_\mathrm{h}\bra{n}_\mathrm{m}\operatorname{Tr}_\mathrm{c}(\rho)\ket{n'}_\mathrm{m}\ket{y}_\mathrm{h}$, where $x,y,k,l\in{0,1}$ (analogous for $\rhn{xy;nn'}{c}$) and $\rho_\mathrm{nn'}=\bra{n}_\mathrm{m}\operatorname{Tr}_\mathrm{h}(\operatorname{Tr}_\mathrm{c}(\rho))\ket{n'}_\mathrm{m}$. For this derivation, we assume $\gh\ll\gc$, allowing us to separate the timescales on which the two ions evolve. The opposite case, $\gh\gg\gc$ is analogous with heating and cooling ion exchanged. We start with the dynamical equations for the levels of the heating ion evaluated at the generic Fock level $n;n'$. Due to the timescale separation, we can ignore the dynamics induced by the cooling ion Hamiltonian (c.f. Eq.~\eqref{twoion:Hc}) 
\begin{align}
    \rhnd{00;nn'}{h}&=-i\gh(\sqrt{n+1}\rhn{10;n+1n'}{h}-\sqrt{n'+1}\rhn{01;nn'+1}{h})\nonumber\\
    &+\gamh\rhn{11;nn'}{h}\label{eq:twoion:qt:sl:dyn1h},\\
    \rhnd{01;nn'+1}{h}&=-i\gh(\sqrt{n+1}\rhn{11;n+1n'+1}{h}-\sqrt{n'+1}\rhn{00;nn'}{h})\nonumber\\
    &-\frac{\gamh}{2}\rhn{01;nn'+1}{h},\\
    \rhnd{10;n+1n'}{h}&=-i\gh(\sqrt{n+1}\rhn{00;nn'}{h}-\sqrt{n'+1}\rhn{11;n+1n'+1}{h})\nonumber\\
    &-\frac{\gamh}{2}\rhn{10;n+1n'}{h},\\
    \rhnd{11;n+1n'+1}{h}&=-i\gh(\sqrt{n+1}\rhn{01;nn'+1}{h}-\sqrt{n'+1}\rhn{10;n+1n'}{h})\nonumber\\
    &-\gamh\rhn{11;n+1n'+1}{h}.\label{eq:twoion:qt:sl:dyn4h}
\end{align}
The full derivation of these equations can be found in Appendix~\ref{sec:append:derivationOfEOM}.

This set of dynamical equations cannot be solved as it does not close, owing to the last term in equation Eq.~\eqref{eq:twoion:qt:sl:dyn1h}. In order to express $\rhn{11;nn'}{h}$ in terms of $\rhn{00;nn'}{h}$ and $\rho_\mathrm{nn'}$, we use the trace over both spin Hilbert spaces: $\rho_\mathrm{nn'}=\rhn{00;nn'}{h}+\rhn{11;nn'}{h}$. The state $\rhn{00;nn'}{h}$ appears on the left-hand side in the set of equations, but not $\rho_\mathrm{nn'}$. We therefore make the assumption that $\rho_\mathrm{nn'}$ remains constant in time. This is based on the assumption that $\expval{n}\gg1$, analogous to the mean-field approach, with the key difference that here we do not have to assume the steady state to be coherent. The now closing set of dynamical equations are integrated, with a constant inhomogeneity given by $\gamh\rho_\mathrm{nn'}$. As specified in Appendix~\ref{sec:append:derivationOfEOM}, we find the following dynamical equations for the heating spin coherence evaluated for $n=n'$
\begin{align}
    \rhn{01;nn+1}{h}&=\frac{2i \gamh \gh \sqrt{1 + n} }{\gamh^2 + 8\gh^2 (1 + n)}\rho_\mathrm{nn},\label{eq:qt:heating:01}\\
    \rhn{10;n+1n}{h}&=-\frac{2i\gamh \gh \sqrt{1 + n}}{\gamh^2 + 8 \gh^2 (1 + n)}\rho_\mathrm{nn}.\label{eq:qt:heating:10}
\end{align}
In the next step, we calculate the dynamics of the cooling ion. For this, we can no longer neglect the contribution of the heating Hamiltonian, as $\gh\ll\gc$. Instead, we use expressions in Eqs.~\eqref{eq:qt:heating:01} and \eqref{eq:qt:heating:10} for the heating ion levels. The detailed derivation can be found in the Appendix \ref{sec:append:derivationOfEOM}. We state the results of the calculation
\begin{align}
    \rhn{01;nn-1}{c}&=\frac{2i \gc \sqrt{n}}{\gamc^2+8\gc^2\gamc^2n} \left[\left(\gamc- \frac{4 \frac{\gh^2}{\gamh}(1 + n)}{1 + 8 \frac{\gh^2}{\gamh^2} (1 + n)}\right) \rhn{nn}{}\nonumber\right.\\
    &+ \left.\frac{4 \frac{\gh^2}{\gamh}n }{1 + 8 \frac{\gh^2}{\gamh^2} n}\rhn{n-1n-1}{} \right].\label{eq:qt:cooling:01}
\end{align}

We have thus found an expression for the spin coherence at a given $n$ in terms of the reduced density matrix $\rho_\mathrm{nn'}$ and the system parameters. Using these results, we simplify the dynamical equation for the motional state of a given level $n,n'$
\begin{align}
    \dot{\rho}_\mathrm{nn'}=&-i\gh(\sqrt{n+1}\rhn{10;n+1n'}{h}-\sqrt{n'}\rhn{10;nn'-1}{h}\nonumber\\
    &+\sqrt{n}\rhn{01;n-1n'}{h}-\sqrt{n'+1}\rhn{01;nn'+1}{h})\nonumber\\
    &-i\gc(\sqrt{n}\rhn{10;n-1n'}{c}-\sqrt{n'+1}\rhn{10;nn'+1}{c}\nonumber\\
    &+\sqrt{n+1}\rhn{01;n+1n'}{c}-\sqrt{n'}\rhn{01;nn'-1}{c}),\label{eq:2ion:mf:pndot}
\end{align}
by using the expressions in Eqs.~\eqref{eq:qt:heating:01} and \eqref{eq:qt:cooling:01} (and the c.c. thereof).
Note that $\rho_\mathrm{nn'}$ corresponds to an element of the density matrix which lies on the $n-n'$ off-diagonal. In the following, we restrict ourselves to the diagonal elements, i.e. $n=n'$, in order to find a recurrence relation for the occupation probability of a given level $n$ 
\begin{align}
    \dot{p}(n)=&f_1(n)p(n-1)-f_1(n+1)p(n)\nonumber\\
    &-f_2(n)p(n)+f_2(n+1)p(n+1),\label{eq:2ion:qt:pndot}
\end{align}
where $f_\mathrm{1,2}(n)$ are functions of heating and cooling spin parameters ($\gh$, $\gc$, $\gamh$, $\gamc$), as well as the levels $n$, and we have used the notation $p(n)=\rhn{nn}{}$. Eq.~\eqref{eq:2ion:qt:pndot} can be split into two equivalent relations: one evaluated at $n$ and a second one with the same terms evaluated at $n+1$. Using this, we find the solution for the relation $f_1(n)p(n-1)-f_2(n)p(n)=0$ which reads
\begin{align}
    p(n)=p(0)\prod_\mathrm{k=1}^\mathrm{n}\left(\frac{f_\mathrm{1}(k)}{f_\mathrm{2}(k)}\right).
\end{align}
The value for $p(0)$ can be determined through the normalization condition $\sum_\mathrm{n}p(n)=1$
\begin{align}
    p(0)=\left(\sum_\mathrm{n=0}^\mathrm{\infty}\prod_\mathrm{k=1}^\mathrm{n}\frac{f_\mathrm{1}(k)}{f_\mathrm{2}(k)}\right)^{-1}.
\end{align}

With this, we found an analytic expression for $p(n)$ which is given in its full form in the Appendix \ref{sec:append:derivationOfG2}. Using this expression for $p(n)$ in the definition of the second-order coherence function given by
\begin{align}
    \gzn=\frac{\expval{a^\dagger a^\dagger a a}}{\expval{a^\dagger a}^2} = \frac{\sum_\mathrm{n}n(n-1)p(n)}{(\sum_\mathrm{n}np(n))^2},\label{eq:qt:slm:g2:as}
\end{align}
we obtain an explicit form of $\gzn$, which is given in the Appendix \ref{sec:append:derivationOfG2}. This result contains hypergeometric functions and is not straightforward to interpret. We therefore make an additional assumption: $\gamh\ll\gamc$, which allows us to Taylor expand in orders of $\gamh/\gamc$. To lowest order, we find the analytic result
\begin{align}
    \gzn = 2-8\frac{\frac{2\gh^2}{\gamh^2}-\frac{\gc^2}{\gamc^2}}{(1+4\frac{\gc^2}{\gamc^2})(1+16\frac{\gh^2}{\gamh^2})}.\label{eq:qt:slm:g2}
\end{align}
For $\gamh/\gamc = 0.5$ and $\gh^2/\gc^2 = 0.125$ (i.e. below the lasing threshold and within the stable regime) we obtain $\gzn=2$, corresponding to a thermal state. This result is in agreement with our previous mean-field analysis (c.f. Section~\ref{sec:twoionPLM}), where we expect to find the phonon laser in the dark state for these parameters. In order to gain more intuition for this result, we investigate it in the limit of $\gc\ll\gamc$. This corresponds to the assumption that the cooling ion transition is heavily damped. In this case, we find the following expression in the lowest two orders of $\gc/\gamc$
\begin{align}
    \gzn \approx \left(1+\frac{1}{1+16\frac{\gh^2}{\gamh^2}}\right)\left(1+4\frac{\gc^2}{\gamc^2}\right).\label{eq:2ion:qt:g2approx}
\end{align}
We first only consider the lowest order $\gc/\gamc$, i.e. neglect the second term in the right bracket. 
If in addition the heating ion transition is heavily damped, i.e. $\gh\ll\gamh$, the effective action of both ions on the motional mode is incoherent. The system is therefore in a thermal state where $\gzn=2$. Depending on whether the heating or cooling process has a stronger effective rate, the phonon laser is either in the heating or in the dark phase. By increasing $\gh$ with respect to $\gamh$ we find $\gzn=1$, which is the hallmark of a coherent state and lasing dynamics. This agrees with the intuition that as the heating process related to $\kaph$ becomes more coherent, we approach a regime where lasing takes place. Note that as soon as the assumption $\gc\ll\gamc$ is relaxed and when taking into account second order terms, $\gzn$ increases and the system moves from a coherent ($\gzn=1$) towards a thermal state ($\gzn=2$). This highlights the importance for one of the ion transitions to be incoherent in order to achieve coherent lasing dynamics.
\begin{figure}[ht]
    \centering
    \includegraphics[width=0.5\textwidth]{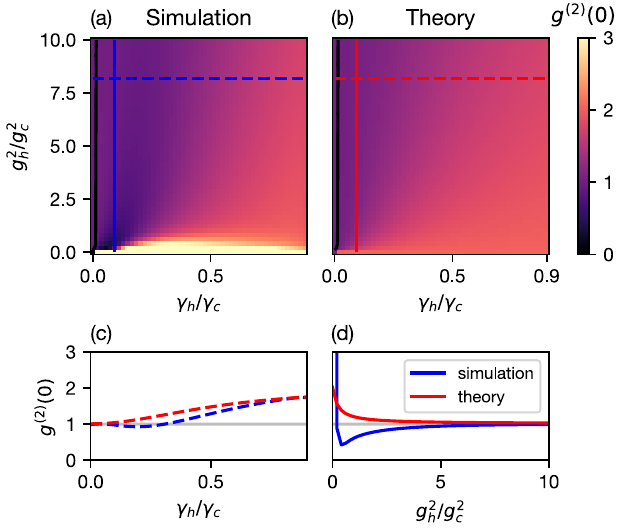}
    \caption{Comparison of the second-order coherence function $\gzn$ between simulation (a) and theory (b). For the simulation we kept both $\gh$ and $\gamh$ fixed (at $4.0$ and $1.5$ respectively). (c) and (d): Two cuts of the parameter space along the blue and red lines in (a) and (b). The crossing point of the two cuts corresponds to the regime, where we expect our theory to be correct, i.e. the used approximations are satisfied. Indeed, in this area agreement of theory and simulation is found. As we leave this area towards larger $\gc$ (down in parameter space) the theory starts to diverge from the simulation. The qualitative behaviour is still given if we move towards lower $\gamc$ (to the right in parameter space). Here, we see the expected transition from coherent to thermal as we move from the lasing to the heating phase.}
    \label{fig:G2Comparison}
\end{figure}

\begin{figure}[ht]
    \centering
    \includegraphics[width=0.5\textwidth]{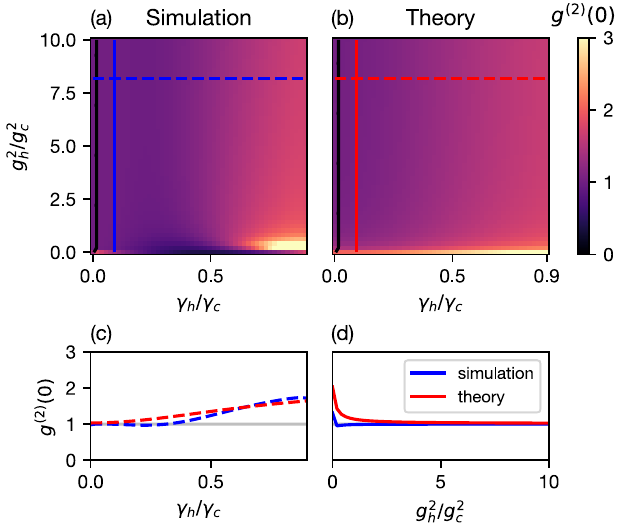}
    \caption{Comparison between the second-order coherence function $\gzn$ between simulation (a) and theory (b). For these plots $\gc$ and $\gamh$ were fixed at $1.0$ and $1.5$, as opposed to $\gh$ and $\gamh$ in Fig.~\ref{fig:G2Comparison}. This allows us to investigate the violation of the assumptions $\gh\ll\gamh$ and $\gc\ll\gh$. (c) and (d): Two cuts of the parameter space along the blue and red lines in (a) and (b). Again we see a good agreement between the two in the parameter space where the approximations made in the derivation are satisfied (crossing of blue/red lines in (a) and (b)). If $\gh$ is decreased we leave the area of coherent lasing and start to violate the assumption $\gc\ll\gh$. Here, the result of the simulation starts to separate from theory, while still showing the same qualitative feature: an increase to $\gzn=2$.}
    \label{fig:G2gcComparison}
\end{figure}

We compare the above result to simulations in Figs.~\ref{fig:G2Comparison} and~\ref{fig:G2gcComparison}. In the former, we keep $\gh$ and $\gamh$ fixed and vary $\gc$ and $\gamc$ along the x and y-axis. While in the latter, we keep $\gc$ and $\gamh$ fixed. In both cases, we see a good match between theory (Eq.~\eqref{eq:qt:slm:g2}) and simulation in the parameter space corresponding to the made assumptions: $\gh\gg\gc$ and $\gamh\ll\gamc$. We further investigate the behaviour of Eq.~\eqref{eq:qt:slm:g2} as we leave this parameter regime. If we increase $\gamh/\gamc$, we increase $\gzn$ smoothly from a 1 to 2, indicating that we approach the thermal state observed in the heating phase. As we decrease $\gh$ in Fig.~\ref{fig:G2Comparison} and approach the lasing threshold, theory and simulation start to diverge: while the theory predicts a thermal state below threshold, the simulation at first tends below 1, indicating sub-Poissonian statistics, however, diverges close to zero. This is due to the low steady state phonon numbers occurring at these parameter values. For $\expval{n}$ below $1$, $\gzn$ approaches zero. However, due to an artefact of the calculation, this value diverges for $\expval{n}\rightarrow0$.

\subsection{Higher-order Lamb-Dicke terms as a saturation mechanism}
The operation of a laser is based on a process which saturates, allowing it to reach a finite steady state. In many lasers, an effective two-level system, which reaches the fully mixed state, provides this saturating effect. This is also the case for the previous analysis, where the Lamb-Dicke approximation is used only to first order. It is however also possible to reach saturation of the lasing process using higher-order terms of the Lamb-Dicke approximation. These can be included when deriving the two sideband Hamiltonian's in Eqs.~\eqref{twoion:Hh} and~\eqref{twoion:Hc}~\cite{Rojkov2024,Simoni2025}. We illustrate this for the Hamiltonian of the heating ion. To first order in the Lamb-Dicke~(LD) expansion, we find the Hamiltonian $H_\mathrm{h}$, given in Eq.~\eqref{twoion:Hh}. However, if we include the third-order term of the Lamb-Dicke expansion, we get the following modified Hamiltonian of the heating ion
\begin{align}
    H_\mathrm{h}&=\gh \left(a^\dagger\Big(1-\frac{\nuh^2}{2}a^\dagger a\Big) \sph + \Big(1-\frac{\nuh^2}{2}a^\dagger a\Big) a \smh\right).\label{eq:ld:Hh1}
\end{align}
Note that $\gh \propto \nuh$ to lowest order in $\nuh$. 
In the first-order LD Hamiltonian the terms proportional to the creation and destruction operators grow with a square root dependence on $n$, the number of phonons. In the third-order terms, the dependence is a more complex function of $n$. Note that, in general, the Lamb-Dicke expansion is given by a Laguerre polynomial. We can use this to engineer the slope of heating and cooling rates at their intersection, which modifies the phonon distribution in the steady state. 

In order to illustrate the modified saturation process, we derive the mean-field result using the Hamiltonian in Eq.~\eqref{eq:ld:Hh1} (analogous for the cooling Hamiltonian). We find the following modified condition for the steady state
\begin{align}
    R_\mathrm{h}:=\frac{\kaph(2-I\nuh^2)}{1+2\frac{\gh^2I}{\gamh^2} (2-I\nuh^2)^2}=\frac{\kapc(2-I\nuc^2)}{1+2\frac{\gc^2I}{\gamc^2}(2-I\nuc^2)^2}=:R_\mathrm{c}.
\end{align}
Note that $I$ is the average phonon number. In the limit of $\nuh\ll1$ and $\nuc\ll1$, we recover the first order Lamb-Dicke result Eq.~\eqref{twoion:master:I}. We define the right hand side to be $R_\mathrm{h}$, the effective heating rate and the left hand side $R_\mathrm{c}$, the effective cooling rate. 

\begin{figure}
    \centering
    \includegraphics[width=0.47\textwidth]{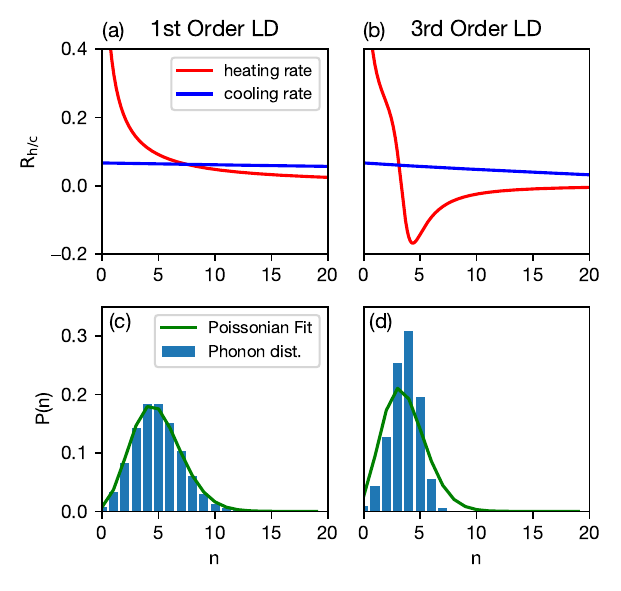}
    \caption{(a) and (b): Comparison of the effective heating and cooling rates $R_\mathrm{h}$ (red) and $R_\mathrm{c}$ (blue) in the first order Lamb-Dicke approximation (left) and to third-order (right) for different phonon occupation $\expval{n}$. The intersection of the effective heating and cooling rates defines the steady-state of the system, the stability of which is determined by the cooling rate exceeding the heating deviations to larger $\expval{n}$ and vice versa. (c) The steady-state phonon distribution (blue bars) according to simulation in the case of first-order Lamb-Dicke approximation. The green line indicates a Poissonian distribution, signifying a coherent state. (d) Simulation of the steady-state phonon distribution of the system including third-order Lamb-Dicke expansion. In this case the occupation deviates from the Poissonian distribution, in line with the steeper intersection of the two rates.}
    \label{fig:LambDickeLayout_half}
\end{figure}

In Fig.~\ref{fig:LambDickeLayout_half}~(a) we show the effective heating and cooling rates $R_\mathrm{h/c}$ without the third-order Lamb-Dicke term, as a function of the phonon number $n$. Their intersection predicts (on a mean-field level) the mean phonon number. The slope of their intersection determines the spread of the phonon distribution. In Fig.~\ref{fig:LambDickeLayout_half}~(c) we see a simulated steady state phonon distribution and a Poissonian fit. We find the system in a coherent state. The addition of the third-order Lamb-Dicke term, allows us to engineer a steeper intersection of the effective heating and cooling rates $R_\mathrm{h}$ and $R_\mathrm{c}$. This in turn causes the phonon distribution to be narrower than in a coherent state, which is reached for the first-order Lamb-Dicke expansion (see Fig.~\ref{fig:LambDickeLayout_half}~(b) and (d)). In other words, we can reach a state which has a sub-Poissonian distribution, i.e. is non-classical. To find an analytic expression for the steady state phonon distribution, one can follow a similar procedure as described in Ref.~\cite{Rojkov2024}. 

We expect the distribution with a lower spread in phonon numbers to be beneficial for sensing protocols, as will be discussed further in the subsection on sensing applications. To engineer the intersection between effective heating and cooling rates, the two Lamb-Dicke parameters $\nuh$ and $\nuc$ can be tuned experimentally. In the case of trapped ions, this tuning can be achieved by adjusting the trap frequency or the angle of the laser drive relative to the ion's motion.

\subsection{Squeezing}\label{subsec:2ion:squ}
\begin{figure}[t!]
    \centering
    \includegraphics[width=0.5\textwidth]{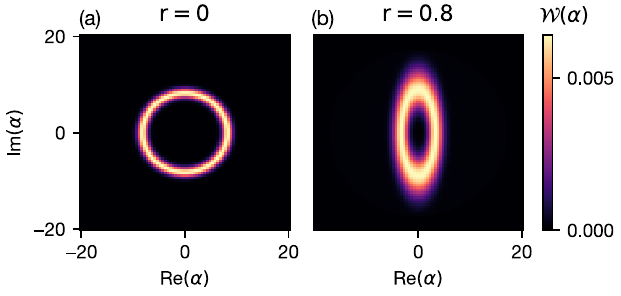}
    \caption{Simulated Wigner function $\mathcal{W}(\alpha)$ for the steady states of the phonon laser, in the phonon field phase space $\alpha$. (a) shows the steady state with a squeezing parameter of $r=0.0$ and (b) with $r=0.8$. As we choose the squeezing phase to be zero ($\beta=0$), we see that the real quadrature of the phonon mode gets squeezed, while the imaginary quadrature is anti-squeezed. This results in decreased variance in the squeezing direction and increased in anti-squeezing direction.}
    \label{fig:SqueezingLayout}
\end{figure}

A further interesting extension of the two-ion lasing model is the addition of lasing in a squeezed basis~\cite{Kienzler2017}. The bosonic squeezing operator is defined as $S(\xi)=e^{(\xi a^2 - \xi (a^\dagger)^2)/2}$, where $\xi=re^{i\beta}$ \cite{Kienzler2017}. In trapped ions, the squeezing of the motional states can be implemented using a bi-chromatic laser drive on both ion species \cite{Kienzler2017,Kienzler2015}. Specifically, both the red and blue motional sidebands must be driven simultaneously on both ions. The Hamiltonians describing the system now read
\begin{align}
    H_\mathrm{h}^\mathrm{BSB}=\gh^\mathrm{BSB}(a^\dagger \sph + a \smh),\label{eq:sq:bsbh}\\
    H_\mathrm{h}^\mathrm{RSB}=\gh^\mathrm{RSB}(a^\dagger \smh + a \sph),\label{eq:sq:rsbh}\\
    H_\mathrm{c}^\mathrm{BSB}=\gc^\mathrm{BSB}(a^\dagger \spc + a \smc),\label{eq:sq:bsbc}\\
    H_\mathrm{c}^\mathrm{RSB}=\gc^\mathrm{RSB}(a^\dagger \smc + a \spc),\label{eq:sq:rsbc}
\end{align}
We add the two Hamiltonians on each of the ions and combine them to a Hamiltonian which is an effective sideband drive for a squeezed mode $A$
\begin{align}
    &H_\mathrm{h:total}=H_\mathrm{h}^\mathrm{BSB}+H_\mathrm{h}^\mathrm{RSB}\\
    &=(\gh^\mathrm{BSB}a^\dagger+\gh^\mathrm{RSB}a)\sph + (\gh^\mathrm{BSB}a+\gh^\mathrm{RSB}a^\dagger)\smh.
\end{align}
By defining $A:=\cosh(r)a+\sinh(r)a^\dagger$, $\gh^\mathrm{BSB}:=\gh \cosh(r)$ and $\gh^\mathrm{RSB}:=\gh \sinh(r)$, which amounts to performing a Bogoliubov transformation, we find the following sideband Hamiltonian with the squeezed motional mode
\begin{align}
    H_\mathrm{h:total}=\gh(A^\dagger \sph + A \smh).\label{eq:hsq}
\end{align}
Analogously, the cooling Hamiltonian in the squeezed basis now reads
\begin{align}
    H_\mathrm{c:total}=\gc(A^\dagger \smc + A \spc).\label{eq:csq}
\end{align}
With this Bogoliubov transformation of our bosonic mode, we found the same lasing dynamics, i.e. the same phase diagram as for the basic phonon lasing model, with the key difference that we now lase in a squeezed mode given by $A$ and $A^\dagger$. In Fig.~\ref{fig:SqueezingLayout} we compare the simulated phase space distribution of a non-squeezed phonon laser steady state (a) and a steady state, where we set the squeezing parameter to $r=0.8$ (b). As expected from our discussion above, the steady state of the squeezed laser shows a similar shape up to the coordinate transformation of the underlying space. Using the scheme outlined above, we can therefore prepare a squeezed lasing state using a phonon laser. We will discuss in the next subsection explicitly how this can improve a phonon-laser based sensing scheme.
\subsubsection{Applications in sensing}
We can use the symmetry breaking of the U(1) symmetric steady state of the phonon laser when subjected to an external periodic drive as an amplitude sensing scheme, as demonstrated in Ref.~\cite{Behrle2023}. The highest susceptibility of the system is obtained near the lasing phase transition~\cite{Porras2017}. 

In order to quantify the sensitivity enhancement of the squeezed lasing scheme, we consider a resonant external signal of the following form
\begin{align}
    H_\mathrm{signal} = \epsilon^* a + \epsilon a^\dagger,
\end{align}
where $\epsilon=\abs{\epsilon}e^{i\phi}$, is the complex amplitude of the signal to be measured.
We now apply squeezing to our phonon laser with the coefficient $\xi=re^{i\beta}$.
By calculating the Fisher information (see Appendix \ref{sec:append:derivationOfFisher}) we find an upper bound for the maximal achievable precision with an ideal measurement. For the phonon laser steady state close to the phase transition we find the modified Fisher information (and thereby the increase in maximal sensitivity) due to the applied squeezing
\begin{align}
    F_Q[\rho_{\abs{\epsilon}}]\approx\frac{2I^2}{A^2}W^2,
\end{align}
where $W=\cosh(2r)-\cosh(r)\sinh(r)\cos(\beta-2\phi)$ is the improvement from squeezing, $I$ the steady state phonon occupation and $A=\gh^2/\gamh$. Note, to reach the maximum sensitivity, the squeezing has to be applied in phase with the applied signal. In this case, we find an improvement in sensitivity $\propto 1/\cosh(2r)^2$, as compared to the non-squeezed case. For an experimentally obtainable squeezing parameter of $r=1.45$ (see Ref.~\cite{Kienzler2015}) this results in an improvement of the sensitivity by a factor of approximately $80$.

\subsubsection{Experimental considerations}
For a standard phonon lasing scheme close to the phase transition sensitivities to an applied signal of up to $\Delta F\approx 53~\mathrm{yN}/\sqrt{\mathrm{I}}$ can be reached~\cite{Porras2017}. This result was calculated for a ${}^{40}\mathrm{Ca}^+$ ion and a trapping frequency of $\omega_T/(2\pi)=10~\mathrm{MHz}$ and $\gh^2/\gamh\approx0.4~\mathrm{kHz}$. In the squeezed lasing case, bi-chromatic drives are required and their individual coupling rates have to be increased. For example, a squeezing parameter of $r=1.45$ requires for the two sideband drives (Eq.~\eqref{eq:sq:bsbh},\eqref{eq:sq:rsbh}): $\gh^\mathrm{RSB}\approx2.01\gh$ and $\gh^\mathrm{BSB}\approx2.25\gh$; and analogous for the two cooling ion sidebands (Eq.~\eqref{eq:sq:bsbc},\eqref{eq:sq:rsbc}). As discussed in \cite{Kienzler2015} an experimental limitation for increasing the squeezing parameter even further is the violation of the Lamb-Dicke limit, which is reached for approx. $r=2.9$ given a Lamb-Dicke parameter of $\eta=0.05$. As discussed in the previous section, the violation of the Lamb-Dicke limit is not detrimental for the lasing dynamics and instead can even be used as a tool to generate non-classical states~\cite{Rojkov2024}.

The squeezing further increases the ground state heating rate by a factor of $\cosh(2r)/2$, which for $r=1.45$ results in $\cosh(2r)/2\approx4.56$. This limits the coherence time of the motional state and therefore places a limit for the duration of the sensing scheme. However, with coherence times on the order of $30\,\text{ms}$ (see Ref.~\cite{Kienzler2015}), this squeezed sensing scheme lies well within current experimental capabilities.

\section{Single-ion phonon lasing model}
\label{sec:singelIonPLM}
In this section we apply the theoretical methods demonstrated in the previous section to a novel phonon lasing model: we investigate the possibility of lasing with a single ion. While the two-ion mixed species approach \cite{Behrle2023} uses the benefits of spectral isolation of the two species to implement independent dissipation channels, it requires the control of two ions and two complex laser systems. These aspects motivated the investigation of a phonon laser using a single ion.

\begin{figure}[ht]
    \centering
    \includegraphics[width=0.5\textwidth]{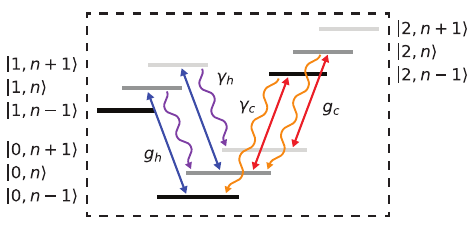}
    \caption{The single-ion lasing scheme is based on three internal levels of the trapped ion. The transition $\ket{0}\leftrightarrow\ket{1}$, is driven with a blue motional sideband with effective strength $\gh$. The $\ket{0}\leftrightarrow\ket{2}$ is driven with a red motional sideband with effective strength $\gc$. The two excited levels $\ket{1}$ and $\ket{2}$ decay with rates $\gamh$ and $\gamc$, respectively.}
    \label{fig:singleIonScheme}
\end{figure}

In the following, we propose a single ion model with three internal states in the resolved sideband regime: $\ket{0}$, $\ket{1}$ and $\ket{2}$. A red motional sideband drive addresses the transition between $\ket{0}$ and $\ket{2}$ and a blue motional sideband drive the transition between $\ket{0}$ and $\ket{1}$, as illustrated in Fig.~\ref{fig:singleIonScheme}. The Hamiltonians implementing these interactions are given by
\begin{align}
    H_\mathrm{h} &= \gh(\ket{2}\bra{0} a^{\dagger} + \ket{0}\bra{2} a),\label{eq:single:mf:Hbsb}\\
    H_\mathrm{c} &= \gc(\ket{1}\bra{0} a + \ket{0}\bra{1} a^{\dagger}).\label{eq:single:mf:Hrsb}
\end{align}

We retain the notation of the two-ion lasing model: $\gh$ and $\gc$ are the coupling strengths of the two sideband drives and subscripts $\mathrm{h}$ and $\mathrm{c}$ refer to blue and red sidebands, respectively.
As in the two-ion lasing model, the two upper levels $\ket{1}$ and $\ket{2}$ are subject to decay to $\ket{0}$ with a time constant $\gamh$ and $\gamc$, respectively. These decays are described with the two jump operators
\begin{align}
    L_\mathrm{h}=\sqrt{\gamh}\ket{0}\bra{2} \quad \text{and} \quad L_\mathrm{c}=\sqrt{\gamc}\ket{0}\bra{1}.\label{eq:singleion:dissipator}
\end{align}
For the following analysis, we define the Hermitian operators
\begin{align}
    \sph&=\ket{2}\bra{0},\spc=\ket{1}\bra{0},\\
    \szh&=\ket{2}\bra{2}-\ket{0}\bra{0},\\
    \szc&=\ket{1}\bra{1}-\ket{0}\bra{0},\\
    \sto&=\ket{2}\bra{1}.
\end{align}

\subsection{Mean-field analysis}
As in the two-ion case, we start with an investigation of the system with a mean-field approach. This will result in a formula for the mean phonon number and allow us to partition the parameter space into different phases. 

We start with the Lindblad master equation in the Heisenberg picture given in Eq.~\eqref{eq:2ion:mf:oeevolution}, which we use to derive the evolution equation for the following observables

\begin{widetext}
\begin{align}
    \frac{d}{dt}\expval{a} =& -i\gc\expval{\smc} -i\gh\expval{\sph},\\
    \frac{d}{dt}\expval{\sph} =&-i\gh\expval{a}\expval{\szh}-i\gc\expval{a^\dagger}\expval{\sto} -\frac{\gamh}{2}\expval{\sph},\\
    \frac{d}{dt}\expval{\spc} =&-i\gc\expval{a^\dagger}\expval{\szc}-i\gh\expval{a}\expval{\sot} -\frac{\gamc}{2}\expval{\spc},\\
    \frac{d}{dt}\expval{\szh} =& 2i\gh\left(\expval{a}\expval{\smh} - \expval{a^\dagger}\expval{\sph}\right) +i\gc\left(\expval{a^\dagger}\expval{\smc} - \expval{a}\expval{\spc}\right)\nonumber\\
    &-\frac{4\gamh}{3}\left(\expval{\szh}+\frac{1}{2}(1-\expval{\szc})\right) -\frac{2\gamc}{3}\left(\expval{\szc}+\frac{1}{2}(1-\expval{\szh})\right),\\
    \frac{d}{dt}\expval{\szc} =& 2i\gc\left(\expval{a^\dagger}\expval{\smc} - \expval{a}\expval{\spc}\right) +i\gh\left(\expval{a}\expval{\smh} - \expval{a^\dagger}\expval{\sph}\right)\nonumber\\
    &-\frac{4\gamc}{3}\left(\expval{\szc}+\frac{1}{2}(1-\expval{\szh})\right) -\frac{2\gamh}{3}\left(\expval{\szh}+\frac{1}{2}(1-\expval{\szc})\right),\\
    \frac{d}{dt}\expval{\sto} = &i\gh\expval{\smc}\expval{a}-i\gc\expval{\sph}\expval{a}-\frac{1}{2}(\gamh+\gamc)\expval{\sto}.
\end{align}
\end{widetext}
Analogous to the two-ion calculation, we can find an evolution equation for $I\approx\expval{a^\dagger}\expval{a}$. By analyzing the roots of the right-hand side of this equation, we find the following steady-state condition for the system
\begin{align}
    I_\mathrm{ss} = \frac{(\gamc + \gamh)(\kaph-\kapc)}{4  \left(\frac{\gc^2}{\gamh} - \frac{\gh^2}{\gamc}\right)(\kaph + \kapc)},\label{eq:nMF:singleIon}
\end{align}
with $\kaph=\frac{\gh^2}{\gamh}$ and $\kapc=\frac{\gc^2}{\gamc}$. The stability of these steady states can be determined by calculating the derivative of the right-hand side with respect to I and evaluating it at the steady state values. This is discussed in more detail in Sec.~\ref{sec:compare}.

\begin{figure}[ht]
    \centering
    \includegraphics[width=0.4\textwidth]{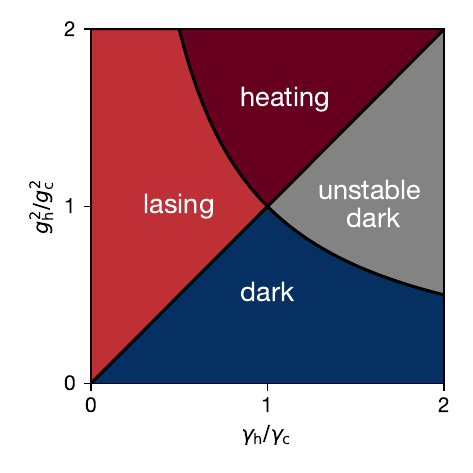}
    \caption{Phase diagram as predicted by the mean-field result Eq.~\eqref{eq:nMF:singleIon}. The y-axis is given by the ratio of the coherent coupling strengths of the JC and AJC Hamiltonian between the internal levels $\ket{0}\leftrightarrow\ket{1}$ and $\ket{0}\leftrightarrow\ket{2}$, respectively. The x-axis is given by the ratio of the incoherent decay rates from the two levels $\ket{1}$ and $\ket{2}$. In the lower area, termed ``dark'', the laser is below the lasing threshold and stable at zero phonons. Crossing the diagonal line, given by $\kaph=\kapc$, we enter the stable lasing phase where a finite steady state is obtained. Above the $1/x$ line, given by $\gh^2/\gc^2=\gamc/\gamh$, no stable solution is found: above the threshold, the phonon number diverges, below a zero steady state exists, which is unstable against perturbations.}
    \label{fig:Single-ion:PD}
\end{figure}

From Eq.~\eqref{eq:nMF:singleIon} we identify the four phases of the single-ion phonon laser, as illustrated in Fig.~\ref{fig:Single-ion:PD}, according to the sign changes of $I_\mathrm{ss}$. The lasing threshold is given by $\kaph=\kapc$. 
The condition for stable solutions is given by $\gh^2/\gc^2<\gamc/\gamh$. In this regime we find a negative phonon number below the lasing threshold ($\kaph<\kapc$). Since in this area the cooling rate exceeds the heating rate, we expect this unphysical result to indicate a steady state of zero phonon occupation, which we term ``dark'' state. Above the lasing threshold, we find a finite stable solution where we expect to find the laser in a coherent state. If we move to the regime of unstable solutions, given by $\gh^2/\gc^2>\gamc/\gamh$, we find a runaway phase above the lasing threshold, where the phonon number diverges (indicated by unphysical negative solution of Eq.~\eqref{eq:nMF:singleIon}, and the heating rate exceeding the cooling rate). Below the lasing threshold, there exists a zero phonon number steady state. We will show later, that this state is unstable against perturbations.

\begin{figure}[ht]
    \centering
    \includegraphics[width=0.5\textwidth]{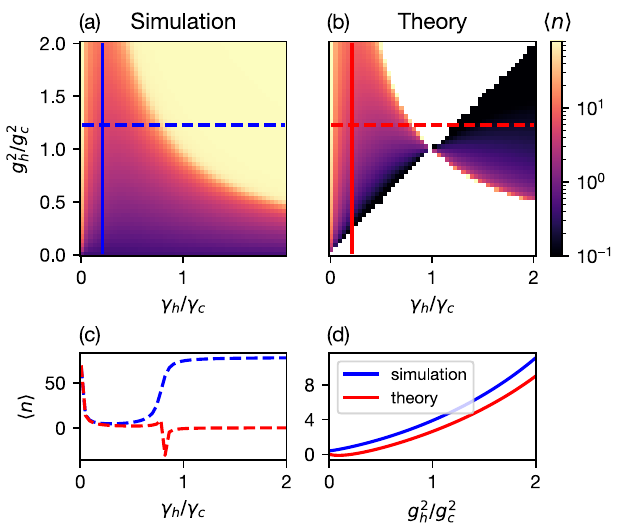}
    \caption{Comparison of the simulated mean phonon number (a) and the derived mean-field theory result (b); both in logarithmic scale. The white shaded areas in the theory plot signify a negative phonon expectation according to Eq.~\eqref{eq:nMF:singleIon}. These solutions are unphysical and instead, following from the stability analysis given in Sec.~\ref{subsec:phasediag}, we expect the heating and dark phase (for the top area and the bottom area, respectively), i.e. zero phonon occupation and a runaway phase. From the mean-field result we see a finite occupation of phonons in the unstable dark area. This phase is not stable against fluctuations and that the phonon number diverges here. (c) and (d): Two cuts of the parameter space along the blue and red lines in (a) and (b).}
    \label{fig:nSingleComparison}
\end{figure}

A comparison of the solution of the mean-field equation Eq.~\eqref{eq:nMF:singleIon} to a simulation of the single ion phonon lasing model is given in Fig.~\ref{fig:nSingleComparison}. The simulation numerically determines steady states of the Liouvillian described by Eqs.~\eqref{eq:single:mf:Hbsb}-\eqref{eq:singleion:dissipator}, and finds a finite solution for the parameter regions corresponding to the dark and lasing phases. In the unstable dark and heating phases, the upper limit of the simulated Fock levels is reached (yellow color). The phonon number smoothly increases as the lasing threshold is crossed from dark to lasing and for very low values of $\gamh/\gamc$ reaches the saturation of the simulation. 
The mean-field theory captures the two transition lines between the four phases as well as the qualitative behaviour in the lasing phase. Note that the white areas signify a negative phonon expectation value according to Eq.~\eqref{eq:nMF:singleIon}. However, as we discuss later, below the lasing threshold the dark state is attained and above it we find a diverging phonon number. The non-zero solution of the center-right area turns out to be unstable against fluctuations and the remaining solution is a runaway phase. This behaviour is reflected in the simulation which reaches saturation.

There are some quantitative discrepancies between theory and simulation: the dark phase, which is predicted to be in the lower area of the phase diagram with zero phonons, is populated with few phonons in the simulation. This kind of discrepancy is however expected from a lowest-order mean-field analysis. To find the phonon distribution of the steady states attained in the lasing and dark phases, we need to investigate the system in a quantum theory. 

\subsection{Quantum theory}
In order to confirm the coherent nature of the steady state in the lasing phase, we have to extend our analysis to a quantum theory including a description of the phonon statistics. We follow a similar procedure as for the two-ion model, where we set up the dynamical equations for the expectation values of individual ion levels. The full derivation is given in the Appendix \ref{sec:append:derivationOfG2Single}.

In the first step, we restrict ourselves to the special case where $\gamh=\gamc$, which is mathematically more tractable. Unlike the general case, this limit allows us to derive an exact expression for the second-order coherence function without further approximations. With this result we confirm a coherent state, however not in the lasing phase, as the lasing phase is present only where $\gamh<\gamc$. The more general case is treated in a second step as an extension of this derivation.
\subsubsection{Special case $\gamh=\gamc$}\label{subsec:G2specialcase}
As described in the Appendix \ref{sec:append:derivationOfG2Single} we find the following evolution equation for the occupation probability $p(n)$
\begin{align}
    \dot{p}(n)&=\frac{\kaph n}{1+8\frac{\gc^2}{\gamma^2}(n-1)+8\frac{\gh^2}{\gamma^2}n}p(n-1)\nonumber\\
    &-\frac{(\kapc n +\kaph (n+1)) }{1+8\frac{\gc^2}{\gamma^2}n+8\frac{\gh^2}{\gamma^2}(n+1)}p(n)\nonumber\\
    &+\frac{\kapc (n+1) }{1+8\frac{\gc^2}{\gamma^2}(n+1)+8\frac{\gh^2}{\gamma^2}(n+2)}p(n+1),
\end{align}
where $\gamma=\gamh=\gamc$.
In order to determine the steady-state phonon distribution, we assume $\dot{p}(n)=0$ and split the above equation into two equivalent recurrence equations. The solution is given by the following steady-state phonon distribution
\begin{align}
    p(n)=p(0)\prod_\mathrm{k=0}^\mathrm{n-1}\left(\frac{\gh^2}{\gc^2}\right)^k\frac{1 + 8\frac{\gc^2}{\gamma^2} k + 8  \frac{\gh^2}{\gamma^2} (1 + k)}{1 + 8 \gh^2}.
\end{align}
The occupation of the vacuum level $p(0)$ is then calculated using the normalization condition $\sum_\mathrm{n}p(n)=1$ and is given by
\begin{align}
    p(0)=\frac{\left(\frac{\gc^2}{\gamma^2} - \frac{\gh^2}{\gamma^2}\right)^2 (1 + 8 \frac{\gh^2}{\gamma^2})}{\frac{\gc^4}{\gamma^4} - \frac{\gc^2 \gh^2}{\gamma^4} + 
 16 \frac{\gc^4 \gh^2}{\gamma^6}}.
\end{align}
With this result for $p(n)$ we use Eq.~\eqref{eq:qt:slm:g2:as} to find the following exact expression for the second-order coherence function
\begin{align}
    \gzn=2 &\frac{(\gamma^2 (\gc^2 - \gh^2)+ 16 \gc^2 \gh^2)}{(\gamma^2 (\gc^2 - \gh^2)+8 (\gc^4 + 3 \gc^2 \gh^2))^2}\times\nonumber\\
   & (\gamma^2 (\gc^2 - \gh^2) + 16 (\gc^4 + 2 \gc^2 \gh^2))\label{eq:singleIon:g2}.
\end{align}

We investigate three interesting limits of Eq.~\eqref{eq:singleIon:g2}: firstly, if $\gc\ll\gamma$
\begin{align}
    \gzn\approx2+\mathcal{O}(\gc^4),
\end{align}
showing that we obtain a thermal state if the transition on the cooling ion is overdamped. In the second case, where $\gh\ll\gamma$, a Taylor expansion yields the following result
\begin{align}
    \gzn\approx2\frac{\left(1+16\frac{\gc^2}{\gamma^2}\right)}{\left(1+8\frac{\gc^2}{\gamma^2}\right)^2}+\mathcal{O}(\gh^2).
\end{align}
In this case, when the heating transition is overdamped, $\gzn$ can reach any value between $0$ and $2$. We therefore find an asymmetry between the heating and cooling coefficients. Lastly, we investigate a limit, where both $\gamma\ll\gh$ and $\gamma\ll\gc$, and find the following expression
\begin{align}
    \gzn\approx\frac{8\gh^2(\gc^2+2\gh^2)}{(\gc^2+3\gh^2)^2}+\mathcal{O}(\gamma^2).
\end{align}
In this limit $\gzn=1$ for a specific ratio of the two coherent driving strengths: $7\gh^2=(2\sqrt{2}-1)\gc^2$.

\begin{figure}[ht]
    \centering
    \includegraphics[width=0.5\textwidth]{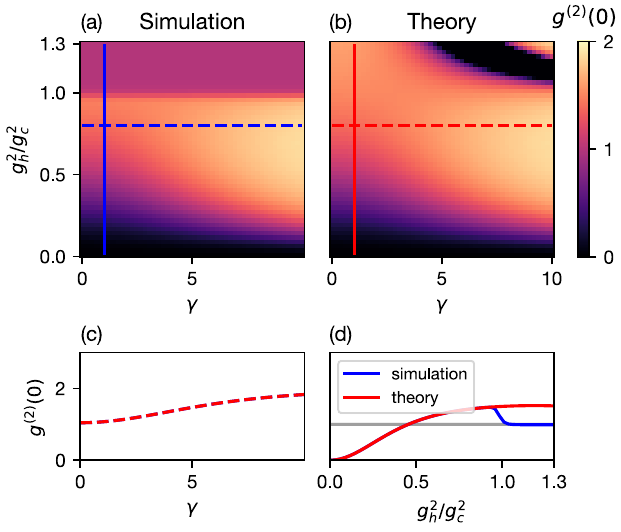}
    \caption{Comparison of the second-order coherence function of simulation (a) and theory (b) derived for the special case $\gamh=\gamc$. As the derivation did not contain any assumptions and simplifications, theory and simulation agree in the finite phonon regime. Note that above $\gh=\gc$, the system is expected to be in the heating phase and the phonon number diverges, we therefore do not expect a physically relevant result from the simulation in this area. (c) and (d): Two cuts of the parameter space along the blue and red lines in (a) and (b).}
    \label{fig:G2SingleGammaComparison}
\end{figure}

The exact expression Eq.~\eqref{eq:singleIon:g2} is compared to a simulation of the single-ion lasing model, and shown in Fig.~\ref{fig:G2SingleGammaComparison}. Due to the exact nature of the result, we see agreement between theory and simulation in the finite phonon number regime. From the phase diagram in Fig.~\ref{fig:Single-ion:PD} we can see that this line does not pass through the lasing phase. We can therefore not confirm the coherent nature of the steady state attained by the system in the lasing phase. For this, we extend this approach to investigate a regime where $\gamh<\gamc$ in the next subsection.

\subsubsection{General case}\label{subsec:G2generalcase}
The above derivation is insightful, as we got a compact formula for $\gzn$. However, this result is only valid for a special hypersurface of the parameter space. To derive an analytic expression for $\gzn$ we assume $\gamh\approx\gamc$. The resulting expression agrees well even when moving further away from $\gamh=\gamc$ into the lasing phase, where $\gamh<\gamc$. The drawback of this general approach is the complexity of the resulting expression containing hypergeometric functions, which cannot easily be simplified. We therefore restrict ourselves to plotting this function and comparing it with the simulated values of $\gzn$. 

\begin{figure}[ht]
    \centering
    \includegraphics[width=0.5\textwidth]{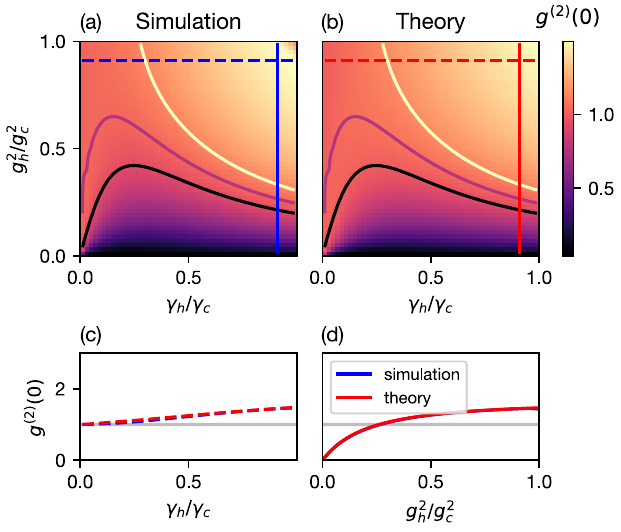}
    \caption{Comparison of the second-order coherence function of simulation (a) and theory (b) derived for the more general case where $\gamh\approx\gamc$. We find excellent agreement between theory and simulation. The two cuts shown in (c) and (d) are taken along the blue and red lines in (a) and (b), in a regime where we expect the above condition to be fulfilled.}
    \label{fig:G2SingleComparison}
\end{figure}

The comparison is shown in Fig.~\ref{fig:G2SingleComparison}. The theory corresponds well to the simulation even when moving away from the area where $\gamh\approx\gamc$. In the lasing phase, we recover values of $\gzn\approx1$. This concludes the quantum theory analysis as we have shown that in the lasing phase we indeed have a coherent state. 

\subsection{Lamb-Dicke terms as saturation mechanism}
As for the two-ion phonon laser, we investigate the effect of higher-order terms of the Lamb-Dicke approximation. Specifically, we derive the mean-field steady state phonon occupation for the single-ion lasing scheme, including the third-order term in the Lamb-Dicke expansion
\begin{align}
    I_\mathrm{ss} = \frac{(\gamc + \gamh)(\kaph-\kapc)}{4  (\frac{\gc^2}{\gamh} - \frac{\gh^2}{\gamc})(\kaph + \kapc)+(\gamc + \gamh)(\kaph\nuh^2-\kapc\nuc^2)}.
\end{align}
For the limit of $\nuh^2\ll1$ and $\nuc^2\ll1$, this reduces to the mean-field result Eq.~\eqref{eq:nMF:singleIon}. The effect of the higher-order terms on the phonon distribution is similar to the two-ion case (c.f. Fig.~\ref{fig:LambDickeLayout_half}). An example of a steady state phonon distribution including third order terms of the LD expansion is given in the Appendix~\ref{sec:append:singleIonSim} and shows similar sub-Poissonian phonon statistics.

\subsection{Squeezing}
We also introduce squeezed lasing dynamics in the single ion model. Analogously to the two-ion model, this is realized by adding a bi-chromatic sideband drive to both the cooling and heating transitions. Specifically, the combination of blue and red motional sideband drive on a single transition can be mapped to a squeezed sideband drive by a Bogoliubov transformation, analogous to subsection \ref{subsec:2ion:squ}. The two resulting Hamiltonians take the form
\begin{align}
    H_\mathrm{h:total}&=\gh(A^\dagger \ket{2}\bra{0} + A \ket{0}\bra{2}),\\
    H_\mathrm{c:total}&=\gc(A^\dagger \ket{0}\bra{1} + A \ket{1}\bra{0}).
\end{align}
This allows us to map the system back onto the single-ion phonon laser Hamiltonian's (Eq.~\eqref{eq:single:mf:Hbsb} and \eqref{eq:single:mf:Hrsb}), with the key difference that we now lase in a squeezed bosonic mode given by $A$ and $A^\dagger$. This mapping and a simulation of the steady state (c.f. Appendix~\ref{sec:append:singleIonSim}) demonstrates that the concept of squeezed lasing is not limited to the two-ion case but can also be realized for a single ion.
\section{Comparison of single and two-ion lasing model}\label{sec:compare}
\label{sec:Comparison}
As explored in the previous sections, the two-ion and single-ion lasing scheme exhibit similar lasing dynamics within certain parameter regimes, but differ in others. In this section, we examine these differences in detail and explore their underlying physical origins.
\subsection{Stability of solutions in the phase diagram}\label{subsec:phasediag}
\begin{figure}[ht]
    \centering
    \includegraphics[width=0.5\textwidth]{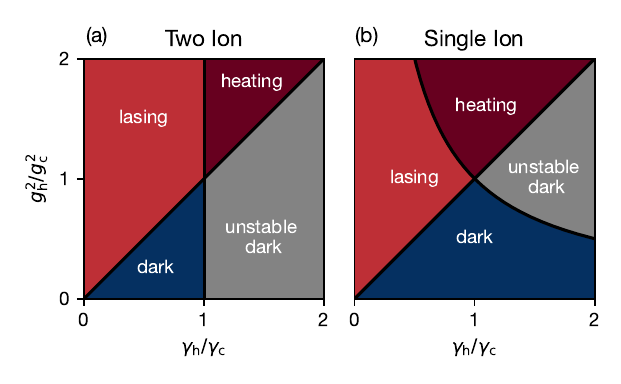}
    \caption{
    Phase diagram of the two-ion and single-ion phonon lasing model, as predicted from a mean-field analysis. Both systems have two transitions: the lasing transition at $\kaph=\kapc$, which is given by the diagonal line and is the same for both models. The second transition is given by the sign change of the non-linear term in the differential equation. Here, the two models differ: for the two-ion laser, the condition for this transition is given by $\gamh=\gamc$. For the single ion model, this transition is a function of the two coherent drive strengths as well as the two dissipation constants.
    Crossing the latter transition to the right causes the non-linear term in the differential equation to be positive. This makes the system generally unbounded. In the unstable dark state, a local minimum exists, but upon crossing the lasing threshold, this minimum disappears and we are left with a diverging system. The four phases remain the same: dark and lasing phase for a negative non-linear term; heating and an unstable dark phase if the non-linear term is positive.}
    \label{fig:phaseDiagBoth}
\end{figure}

As can be seen from Eqs.~\eqref{eq:2ion:mf:n} and~\eqref{eq:nMF:singleIon}, the two models have different phase diagrams and therefore different phase transitions. A schematic depiction of these phase diagrams is given in Fig.~\ref{fig:phaseDiagBoth}.

Let us start by performing a stability analysis of the two-ion model. As discussed, there are two phase transitions one at $\gamh=\gamc$ and one at $\kaph=\kapc$. In order to understand each phase in more detail, we can rewrite Eq.~\eqref{twoion:master:I} into the following form
\begin{align}
    \frac{d}{dt}I\propto32I^2(\gamh-\gamc)+4I\frac{\gamh^2\gamc^2}{\gh^2\gc^2}\left(\kaph-\kapc\right).
\end{align}
Neglecting the non-zero and positive denominator, we analyse the stability of different steady states by investigating the right-hand side of the above equation. Specifically, there are four possible scenarios: Firstly, a stable solution is found at $I_\mathrm{ss}=0$, in the case where the laser is below the lasing threshold ($\kaph<\kapc$) and the non-linear term proportional to $I^2$ is negative ($\gamh<\gamc$). Secondly, there is a lasing phase, which is characterized by the stable solution at a non-zero $I_\mathrm{ss}$. This situation occurs if the laser is above threshold $\kaph>\kapc$ and the non-linear term is negative $\gamh<\gamc$. The second solution $I_\mathrm{ss}=0$ is unstable in this case. Third, the heating phase is given if the laser is above threshold and the non-linear term is positive $\gamh>\gamc$. In this case, an unphysical but stable solution appears at negative $I$. The physical solution at $I_\mathrm{ss}=0$ is unstable and the system therefore diverges. Lastly, the fourth case is given by the laser below threshold with the non-linear term positive $\gamh>\gamc$, therefore allowing the system to diverge if the laser intensity reaches a large enough value. This is reflected by the stable solution at $I_\mathrm{ss}=0$ and the unstable solution at a positive value of $I$. According to mean-field, the system would remain at the stable zero solution. However, taking fluctuations into account, it is plausible that the system diverges in this phase for large enough fluctuations. This behaviour is reflected in simulations. 

We find a similar situation when analysing the single-ion model. Here, in contrast, the non-linear term is given by a combination of $g$ and $\gamma$. Intuitively, this can be understood as the two transitions influencing one another trough the common ground state. The threshold condition is the same as in the two-ion scheme: $\kaph=\kapc$. Again, we investigate the stability of the solutions in the individual phases using the non-linear differential equation for $I$. For simplicity, we exclude the denominator from the description as it is strictly positive for all parameters
\begin{align}
    \frac{d}{dt}I\propto&8I^2\gamh^2\gamc^2\left(\kaph+\kapc\right)\left(\frac{\gh^2}{\gamc}-\frac{\gc^2}{\gamh}\right)\nonumber\\
    +&2I\gamh^2\gamc^2(\gamh+\gamc)\left(\kaph-\kapc\right).
\end{align}

We are interested in the two brackets which change the sign of the terms proportional to $I^2$ and $I$. The dark state is given by the laser being below threshold, i.e. $\kaph<\kapc$ and with negative sign of the non-linear term: $\frac{\gh^2}{\gamc}<\frac{\gc^2}{\gamh}$. Here, there is one stable solution at $I_\mathrm{ss}=0$ and an unphysical, unstable solution at negative intensity. The stable solution becomes unstable when the lasing threshold is crossed but the non-linear term remains negative. Here, a stable solution at a strictly positive value of $I$ appears. This corresponds to the lasing phase. If the non-linear term becomes positive, i.e. $\frac{\gh^2}{\gamc}>\frac{\gc^2}{\gamh}$ and the laser is above threshold, the only stable solution is at an unphysical negative value of $I$, and the solution of $I_\mathrm{ss}$ is unstable. The system therefore diverges. The last phase is below threshold $\kaph<\kapc$ and at positive non-linear term. Here, the mean-field results predict a zero-intensity solution. However, due to the proximity of an unstable solution at a positive value $I$, the system still diverges if fluctuations are large compared to the potential barrier separating the stable from the unstable solution.

This analysis shows that the same four phases are present in both lasing models: dark, lasing, heating with the dark state susceptible to fluctuations. The threshold condition $\kaph=\kapc$ is also the same for both models. The difference appears in the non-linear term, where for the two-ion model this is only given by the relative value of $\gamh$ and $\gamc$. In the single ion model, it is given by a mixture of the $g$ and $\gamma$ parameters of the two transitions. This manifests in a different shape of the phase diagram, as can be seen in Fig.~\ref{fig:phaseDiagBoth}.

\section{Conclusion and outlook}
In this article, we have detailed and extended the existing theory framework of the two-ion phonon laser~\cite{Behrle2023} to a full quantum theory and studied different modifications of this model. With the quantum theory, we were able to find an analytic expression for the second-order coherence function $\gzn$ and theoretically confirm the lasing nature of the steady state above the lasing threshold as observed in experiments \cite{Behrle2023}.

Additionally, we investigated possible extensions of the system, which allow lasing in a squeezed basis as well as obtaining non-classical steady states by utilizing higher-order Lamb-Dicke terms for saturation. We provided a brief discussion on how squeezed states can enhance the sensitivity of phonon laser-based sensing protocols, motivated by the interest in using such systems to perform precision measurements. In experimentally realistic scenarios, our results promise an improvement of sensitivity by a factor of up to 80. 

In the second part, we propose a novel lasing scheme implemented on a single ion. Such a model can significantly reduce experimental efforts, simplifying the implementation of multiple such phonon lasers within the same setup. We performed a mean-field analysis to study the phase diagram and also investigated the model using a quantum theory. The latter allowed us to confirm the lasing nature of the state above threshold. As with the two-ion model, we also explored possible extensions for this system and found that squeezing can be implemented analogously. 

This study paves the way for investigating multiple phonon lasers operating simultaneously in a single setup, both theoretically and experimentally. Such a setting could be interesting for synchronization and sensing experiments. The addition of squeezing to the phonon laser holds the promise to be beneficial for its sensing capabilities. 
More recently, studies of quantum systems similar those from Ref.~\cite{Behrle2023} show the emergence of new interesting physics when going beyond the first-order Lamb-Dicke approximation, such as stabilization of a Schrödinger cat-like state~\cite{Rojkov2024,Simoni2025}. This underscores the rich potential of phonon lasing for advancing both fundamental physics, practical sensing applications as well as new schemes to implement quantum error correction.

\section{Acknowledgments}
We acknowledge support from the Swiss National Science Foundation (SNSF) under grant number 200020 179147/1, the 
SNSF through the National Centre of Competence in Research for Quantum Science and Technology (QSIT) grant 51NF40–185902, the SNFS Ambizione grant PZ00P2-186040 and the EU Quantum Flagship H2020-FETFLAG-2018-03 grant 820495 AQTION. SFY would like to acknowledge funding from the NSF through the CUA PFC.

\bibliography{references}

\onecolumngrid
\newpage
\appendix
\section{Derivation of mean-field equations}
In this section, we derive the time evolution equation for operator expectation values as it was used in the main text in equation Eq.~\eqref{eq:2ion:mf:oeevolution}. The time evolution of an open system, which is described by the density operator $\rho$, is governed by a master equation in Lindblad form
\begin{align}
    \frac{d \rho}{dt} = \mathcal{L}(\rho) = -i [H, \rho] + \sum_k \mathcal{D}[L_k](\rho).\label{eq:master}
\end{align}
The Liouvillian $\mathcal{L}(\rho)$ contains a Hamiltonian part with $H = H_\mathrm{c} + H_\mathrm{h}$, and a dissipator
\begin{align}
    \mathcal{D}[L_k](\rho) = L_k \rho L_k^\dagger - \frac{1}{2}(L_k^\dagger L_k \rho + \rho L_k^\dagger L_k),
\end{align}
for each jump operator $L_k\in\{L_\mathrm{h},L_\mathrm{c}\}$. In the Heisenberg picture, the time evolution of the expectation value of a time-independent operator $O$ is described by
\begin{align}
    \frac{d}{dt} \expval{O} = \frac{d}{dt} \mathrm{Tr}(\rho O) = \mathrm{Tr}\left(\frac{d \rho}{dt} O\right).
\end{align}
With Eq.~\eqref{eq:master} this can be written as
\begin{align}
    \frac{d}{dt} \expval{O} = i \expval{[H,O]} + \sum_k \expval{\tilde{\mathcal{D}}[L_k](O)},
\end{align}
where the dissipative part is given by
\begin{align}
    \expval{\tilde{\mathcal{D}}[L_k](O)} = \expval{L_k^\dagger O L_k - \frac{1}{2}(L_k^\dagger L_k O + O L_k^\dagger L_k)}.
\end{align}
\section{Derivation of equation of motion of the motional state of the two-ion phonon laser model}\label{sec:append:derivationOfEOM}
In this section we provide a detailed derivation of equation Eq.~\eqref{eq:2ion:qt:pndot}: the dynamic equation of the diagonal elements of the reduced density matrix describing the motional mode of the ions. This equation will be the starting point for the derivation of an expression of $\gzn$. In this derivation, we will make use of the separation of the timescale of the two ions given by the expression: $\gh\gg\gc$. We therefore start with the analysis of the heating ion and use the found results in the derivation of the cooling ion.
\subsection{Derivation of the heating spin expressions}
We start by defining the used notation. The state $\rho$ describes the motional state as well as the two spins which are given by heating and cooling ion. Specific levels of the heating and cooling spin at different Fock levels are denoted as $\rho_\mathrm{xy,kl;nn'}:=\bra{x}_\mathrm{h}\bra{k}_\mathrm{c}\bra{n}_\mathrm{m}\rho\ket{n'}_\mathrm{m}\ket{l}_\mathrm{c}\ket{y}_\mathrm{h}$, where $x,y,k,l\in{0,1}$. If we want to talk about the spin state of only one of the ions with the other traced out we write: $\rhn{xy;nn'}{h}:=\bra{x}_\mathrm{h}\bra{n}_\mathrm{m}\operatorname{Tr}_\mathrm{c}(\rho)\ket{n'}_\mathrm{m}\ket{y}_\mathrm{h}$ (analogous for $\rhn{xy;nn'}{c}$). Finally, a Fock level of only the motional mode is given as $\rho_\mathrm{nn'}=\bra{n}_\mathrm{m}\operatorname{Tr}_\mathrm{h}(\operatorname{Tr}_\mathrm{c}(\rho))\ket{n'}_\mathrm{m}$.
With this, we write down the evolution equation of the heating spin levels using the Lindblad master equation.
\begin{align}
    \rhnd{00;nn'}{h}=&\rhnd{00,00;nn'}{}+\rhnd{00,11;nn'}{}\\
    =&-i\gh(\sqrt{n+1}\rhn{10,00;n+1n'}{}-\sqrt{n'+1}\rhn{01,00;nn'+1}{})+\gamh\rhn{11,00;nn'}{}\nonumber\\
    &-i\gh(\sqrt{n+1}\rhn{10,11;n+1n'}{}-\sqrt{n'+1}\rhn{01,11;nn'+1}{})+\gamh\rhn{11,11;nn'}{}\nonumber\\
    &-i\gc(\sqrt{n}\rhn{00,10;n-1n'}{}-\sqrt{n'}\rhn{00,01;nn'-1}{})+\gamc\rhn{00,11;nn'}{}\nonumber\\
    &-i\gc(\sqrt{n+1}\rhn{00,01;n+1n'}{}-\sqrt{n'+1}\rhn{00,10;nn'+1}{})-\gamc\rhn{00,11;nn'}{}\\
    \approx&-i\gh(\sqrt{n+1}\rhn{10;n+1n'}{h}-\sqrt{n'+1}\rhn{01;nn'+1}{h})+\gamh\rhn{11;nn'}{h}\label{eq:append:eom:motion:00}
\end{align}

The first two equations simply use the definition of the notation we have introduced. In the last equation which holds approximately we have neglected the contributions from the cooling ion Hamiltonian, i.e. terms proportional to $\gc$. We perform the analogous calculation for the three remaining heating spin levels:

\begin{align}
    \rhnd{01;nn'+1}{h}=&\rhnd{01,00;nn'+1}{}+\rhnd{01,11;nn'+1}{}\\
    =&-i\gh(\sqrt{n+1}\rhn{11,00;n+1n'+1}{}-\sqrt{n'+1}\rhn{00,00;nn'}{})-\frac{\gamh}{2}\rhn{01,00;nn'+1}{}\nonumber\\
    &-i\gh(\sqrt{n+1}\rhn{11,11;n+1n'+1}{}-\sqrt{n'+1}\rhn{00,11;nn'}{})-\frac{\gamh}{2}\rhn{01,11;nn'+1}{}\nonumber\\
    &-i\gc(\sqrt{n}\rhn{01,10;n-1n'+1}{}-\sqrt{n'+1}\rhn{01,01;nn'}{})+\gamc\rhn{01,11;nn'+1}{}\nonumber\\
    &-i\gc(\sqrt{n+1}\rhn{01,01;n+1n'+1}{}-\sqrt{n'+2}\rhn{01,10;nn'+2}{})-\gamc\rhn{01,11;nn'+1}{}\\
    \approx&-i\gh(\sqrt{n+1}\rhn{11;n+1n'+1}{h}-\sqrt{n'+1}\rhn{00;nn'}{h})-\frac{\gamh}{2}\rhn{01;nn'+1}{h}\label{eq:append:eom:motion:01}
\end{align}

\begin{align}
    \rhnd{10;n+1n'}{h}=&\rhnd{10,00;n+1n'}{}+\rhnd{10,11;n+1n'}{}\\
    =&-i\gh(\sqrt{n+1}\rhn{00,00;nn'}{}-\sqrt{n'+1}\rhn{11,00;n+1n'+1}{})-\frac{\gamh}{2}\rhn{11,00;nn'}{}\nonumber\\
    &-i\gh(\sqrt{n+1}\rhn{00,11;nn'}{}-\sqrt{n'+1}\rhn{11,11;nn'+1}{})-\frac{\gamh}{2}\rhn{11,11;nn'}{}\nonumber\\
    &-i\gc(\sqrt{n+1}\rhn{10,10;nn'-1}{}-\sqrt{n'}\rhn{10,01;n+1n'-1}{})+\gamc\rhn{10,11;n+1n'}{}\nonumber\\
    &-i\gc(\sqrt{n+2}\rhn{10,01;n+2n'}{}-\sqrt{n'+1}\rhn{10,10;n+1n'+1}{})-\gamc\rhn{10,11;nn'}{}\\
    \approx&-i\gh(\sqrt{n+1}\rhn{00;nn'}{h}-\sqrt{n'+1}\rhn{11;n+1n'+1}{h})-\frac{\gamh}{2}\rhn{10;n+1n'}{h}\label{eq:append:eom:motion:10}
\end{align}

\begin{align}
    \rhnd{11;n+1n'+1}{h}=&\rhnd{00,00;n+1n'+1}{}+\rhnd{00,11;n+1n'+1}{}\\
    =&-i\gh(\sqrt{n+1}\rhn{01,00;nn'+1}{}-\sqrt{n'+1}\rhn{10,00;n+1n'}{})+\gamh\rhn{11,00;nn'}{}\nonumber\\
    &-i\gh(\sqrt{n+1}\rhn{01,11;nn'+1}{}-\sqrt{n'+1}\rhn{10,11;n+1n'}{})+\gamh\rhn{11,11;nn'}{}\nonumber\\
    &-i\gc(\sqrt{n+1}\rhn{11,10;n-1n'}{}-\sqrt{n'+1}\rhn{11,01;nn'-1}{})+\gamc\rhn{11,11;nn'}{}\nonumber\\
    &-i\gc(\sqrt{n+2}\rhn{11,01;n+2n'+1}{}-\sqrt{n'+2}\rhn{11,10;n+1n'+2}{})-\gamc\rhn{11,11;nn'}{}\\
    \approx&-i\gh(\sqrt{n+1}\rhn{01;nn'+1}{h}-\sqrt{n'+1}\rhn{10;n+1n'}{h})+\gamh\rhn{11;n+1n'+1}{h}\label{eq:append:eom:motion:11}
\end{align}
The equations Eq.~\eqref{eq:append:eom:motion:00}, Eq.~\eqref{eq:append:eom:motion:01},Eq.~\eqref{eq:append:eom:motion:10} and Eq.~\eqref{eq:append:eom:motion:11} form a coupled set of differential equations. However, this set does not close yet due to the last term in equation Eq.~\eqref{eq:append:eom:motion:00} on the right-hand side. We use the definition of the trace: $\rho_\mathrm{nn'}=\rhn{00;nn'}{h}+\rhn{11;nn'}{h}$ and substitute this definition solved for $\rhn{11;nn'}{h}$. With this, we have found a closing set of differential equations apart from the introduced inharmonicity $\rho_\mathrm{nn'}$. By assuming that this inharmonicity term is constant in time we can still solve the set of equations. This is justified as explained in the main text, since the state we investigate is at a large $\expval{n}$ and has a large variance $\expval{\Delta n}$. Therefore the evolution of the ion does not change the motional state significantly for short times.

Explicitly we solve the set of equations by recasting them into matrix form:
\[
  R(t) =
  \left[ {\begin{array}{c}
    \rhn{00;nn'}{h}(t) \\
    \rhn{01;nn'+1}{h}(t) \\
    \rhn{10;n+1n'}{h}(t) \\
    \rhn{11;n+1n'+1}{h}(t) \\
  \end{array} } \right]
\]
\[
  M =
  \left[ {\begin{array}{cccc}
    \gamh & -i\gh\sqrt{n'+1} & i\gh\sqrt{n+1} & 0 \\
    -i\gh\sqrt{n'+1} & \frac{\gamh}{2} & 0 & i\gh\sqrt{n+1}\\
    i\gh\sqrt{n+1} & 0 & \frac{\gamh}{2} & -i\gh\sqrt{n'+1} \\
    0 & i\gh\sqrt{n+1} & -i\gh\sqrt{n'+1} & \gamh \\
  \end{array} } \right]
\]
\[
  A(t) =
  \left[ {\begin{array}{c}
    \gamh\rhn{nn'}{} \\
    0 \\
    0 \\
    0  \\
  \end{array} } \right]
\]

The equation we want to solve takes the following form:

\begin{align}
    \dot{R}(t) = -MR(t) + A(t). \label{eq:app:3ll:m1:matrixDGL}
\end{align}

By making the time scale separation as explained before, the term $A$ can be assumed to be constant for the evolution of the spin, i.e. for integrating equation Eq.~\eqref{eq:app:3ll:m1:matrixDGL}. The formal solution of equation Eq.~\eqref{eq:app:3ll:m1:matrixDGL} is given by:

\begin{align}
    R(t) = \int_\mathrm{-\infty}^{t} e^{-M(t-t^{\prime})}Adt^{\prime} = M^{-1}A.\label{eq:app:3ll:m1:matrixDGL:solution}
\end{align}

Explicitly this gives:

\[
\begin{aligned}
  R(t) =&
  \left[ {\begin{array}{c}
    \rhn{00;nn'}{h}(t) \\
    \rhn{01;nn'+1}{h}(t) \\
    \rhn{10;n+1n'}{h}(t) \\
    \rhn{11;n+1n'+1}{h}(t) \\
  \end{array} } \right]\\
  &= \frac{1}{\gamh^4 + 4 \gh^4 (n - n')^2 + 4 \gamh^2 \gh^2 (2 + n + n')}
   \left[ {\begin{array}{c}
   \gamh^3 + 2 \gamh \gh^2 (2 + n + n')\\
   -2 i \gh \sqrt{1 + n'} (\gamh^2 + 2 \gh^2 (n' - n))\\
   -2 i \gh \sqrt{1 + n} (\gamh^2 + 2 \gh^2 (n - n')) \\
    4 \gamh \gh^2 \sqrt{1 + n} \sqrt{1 + n'}\\
  \end{array} } \right]
  \gamh \rhn{nn'}{}.
\end{aligned}
\]
With this we have found an expression for the spin levels of the heating ion (Eq.~\eqref{eq:qt:cooling:01} in the main text) in terms of the partial trace $\rhn{nn'}{}$. In the following subsection, we derive an expression for the cooling spin.

\subsection{Derivation of the cooling spin expressions}
For the cooling spin, we have to take the coherent action of the heating ion into account, as $\gh\gg\gc$. To find an expression for the cooling spins we use the solution of the heating ions derived before. However, for this a further approximation is necessary. We assume that the cooling ion predominantly occupies its ground state. In terms of parameters, this condition reads $\gh\ll\gamh$. Using this we approximate the traced-out version of the heating spin levels with the following: $\rhn{xy;nn'}{h}=\rhn{xy,00;nn'}{}+\rhn{xy,11;nn'}{}\approx\rhn{xy,00;nn'}{}$. We further assume the two coherences of the cooling ion to be negligible: $\rhn{xy,01;nn'}{}\approx0$ and $\rhn{xy,10;nn'}{}\approx0$ (where $x,y\in0,1$). This is a result of the approximation, that the cooling spin transition is overdamped.

Lastly, as we are only interested in expressions for the diagonal Fock states we restrict ourselves to the case $n=n'$ in the following in favour of readability. Using the Lindbald master equation, we find the following evolution equations for the cooling spin levels:
\begin{align}
    \rhnd{00;nn}{c}=&\rhnd{00,00;nn}{}+\rhnd{11,00;nn}{}\\
    =&-i\gh(\sqrt{n+1}\rhn{10,00;n+1n}{}-\sqrt{n+1}\rhn{01,00;nn+1}{})+\gamh\rhn{11,00;nn}{}\nonumber\\
    &-i\gh(\sqrt{n}\rhn{01,00;n-1n}{}-\sqrt{n}\rhn{10,00;nn-1}{})-\gamh\rhn{11,00;nn}{}\nonumber\\
    &-i\gc(\sqrt{n}\rhn{00,10;n-1n}{}-\sqrt{n}\rhn{00,01;nn-1}{})+\gamc\rhn{00,11;nn}{}\nonumber\\
    &-i\gc(\sqrt{n}\rhn{11,10;n-1n}{}-\sqrt{n}\rhn{11,01;nn-1}{})+\gamc\rhn{11,11;nn}{}\\
    \approx&-i\gh(2\sqrt{n+1}\rhn{10;n+1,n}{h}+2\sqrt{n}\rhn{01;n,n-1}{h})\nonumber\\
    &-i\gc(\sqrt{n}\rhn{10;n-1n}{c}-\sqrt{n}\rhn{01;nn-1}{c})+\gamc\rhn{11;nn}{c}\\
    =&\left(\gamc-\frac{4\frac{\gh^2}{\gamh}(n+1)}{1+8\frac{\gh^2}{\gamh^2}(n+1)}\right)\rhn{nn}{}+\frac{4\frac{\gh^2}{\gamh}n}{1+8\frac{\gh^2}{\gamh^2}n}\rhn{n-1n-1}{}\nonumber\\
    &-i\gc(\sqrt{n}\rhn{10;n-1n}{c}-\sqrt{n}\rhn{01;nn-1}{c})-\gamc\rhn{00;nn}{c}\label{eq:append:eom:motion:00:c}
\end{align}

Where we have used the definition of our notation in the first two equations. The third equality is given only approximately as we make use of the assumption $\gc\ll\gamc$ in this step. In the last line, we have used the found expression for $\rhn{10;n+1,n}{h}$ and $\rhn{01;n,n-1}{h}$. We find analogous expressions for the other levels in a similar fashion, with the difference, that for the remaining levels, we will assume the contribution of the heating spin to be zero (again due to the assumption $\gc\ll\gamc$). 

\begin{align}
    \rhnd{01;nn-1}{c}=&\rhnd{00,01;nn-1}{}+\rhnd{11,01;nn-1}{}\\
    =&-i\gh(\sqrt{n+1}\rhn{10,01;n+1n-1}{}-\sqrt{n}\rhn{01,01;nn}{})+\gamh\rhn{11,01;nn-1}{}\nonumber\\
    &-i\gh(\sqrt{n}\rhn{01,01;n-1n-1}{}-\sqrt{n-1}\rhn{10,01;nn-2}{})-\gamh\rhn{11,01;nn-1}{}\nonumber\\
    &-i\gc(\sqrt{n}\rhn{00,11;n-1n-1}{}-\sqrt{n}\rhn{00,00;nn}{})+\frac{\gamc}{2}\rhn{00,01;nn-1}{}\nonumber\\
    &-i\gc(\sqrt{n}\rhn{11,11;n-1n-1}{}-\sqrt{n}\rhn{11,00;nn}{})-\frac{\gamc}{2}\rhn{11,01;nn-1}{}\\
    \approx&-i\gh(\sqrt{n}\rhn{11;n-1n-1}{c}-\sqrt{n}\rhn{00;nn}{c})-\frac{\gamh}{2}\rhn{01;nn-1}{c}\label{eq:append:eom:motion:01:c}
\end{align}

\begin{align}
    \rhnd{10;n+1n}{c}=&\rhnd{00,10;n-1n}{}+\rhnd{11,10;n-1n}{}\\
    =&-i\gh(\sqrt{n}\rhn{10,10;nn}{}-\sqrt{n+1}\rhn{01,10;n-1n+1}{})-\gamh\rhn{11,10;n-1n}{}\nonumber\\
    &-i\gh(\sqrt{n-1}\rhn{01,10;n-2n}{}-\sqrt{n}\rhn{10,10;n-1n}{})-\gamh\rhn{11,10;n-1n}{}\nonumber\\
    &-i\gc(\sqrt{n}\rhn{00,00;nn}{}-\sqrt{n}\rhn{00,11;n-1n-1}{})+\frac{\gamc}{2}\rhn{00,10;n-1n}{}\nonumber\\
    &-i\gc(\sqrt{n}\rhn{11,00;nn}{}-\sqrt{n}\rhn{11,11;n-1n-1}{})-\frac{\gamc}{2}\rhn{11,10;n-1n}{}\\
    \approx&-i\gh(\sqrt{n}\rhn{00;nn}{c}-\sqrt{n}\rhn{11;n-1n-1}{c})-\frac{\gamh}{2}\rhn{10;n-1n}{c}\label{eq:append:eom:motion:10:c}
\end{align}

\begin{align}
    \rhnd{11;n-1n-1}{c}=&\rhnd{00,11;n-1n-1}{}+\rhnd{11,11;n-1n-1}{}\\
    =&-i\gh(\sqrt{n}\rhn{10,11;nn-1}{}-\sqrt{n}\rhn{01,11;n-1n}{})+\gamh\rhn{11,11;n-1n-1}{}\nonumber\\
    &-i\gh(\sqrt{n-1}\rhn{01,11;n-2n-1}{}-\sqrt{n-1}\rhn{10,11;n-1n-2}{})-\gamh\rhn{11,11;n-1n-1}{}\nonumber\\
    &-i\gc(\sqrt{n}\rhn{00,01;nn-1}{}-\sqrt{n}\rhn{01,10;n-1n}{})-\gamc\rhn{00,11;n-1n-1}{}\nonumber\\
    &-i\gc(\sqrt{n}\rhn{11,01;nn-1}{}-\sqrt{n}\rhn{11,10;n-1n}{})-\gamc\rhn{11,11;n-1n-1}{}\\
    \approx&-i\gh(\sqrt{n}\rhn{01;nn-1}{c}-\sqrt{n+1}\rhn{10;n-1n}{c})+\gamh\rhn{11;n-1n-1}{c}\label{eq:append:eom:motion:11:c}
\end{align}

With equations Eq.~\eqref{eq:append:eom:motion:00:c}, Eq.~\eqref{eq:append:eom:motion:01:c}, Eq.~\eqref{eq:append:eom:motion:10:c} and Eq.~\eqref{eq:append:eom:motion:11:c} we have found another closing set of coupled differential equations, with a constant in time inharmonicity. We therefore use the same procedure as for the heating spin expressions to write these equations as a matrix and find the solution of this set of equations:

\begin{align}
    \rhn{01;nn-1}{c}&=\frac{2i \gc \sqrt{n}}{\gamc^2+8\gc^2\gamc^2n} \left(\left(\gamc- \frac{4 \frac{\gh^2}{\gamh}(1 + n)}{1 + 8 \frac{\gh^2}{\gamh^2} (1 + n)}\right) \rhn{nn}{} + \frac{4 \frac{\gh^2}{\gamh}n }{1 + 8 \frac{\gh^2}{\gamh^2} n}\rhn{n-1n-1}{} \right)\\
    \rhn{10;n-1n}{c}&=\frac{-2i \gc \sqrt{n}}{\gamc^2+8\gc^2\gamc^2n} \left(\left(\gamc- \frac{4 \frac{\gh^2}{\gamh}(1 + n)}{1 + 8 \frac{\gh^2}{\gamh^2} (1 + n)}\right) \rhn{nn}{} + \frac{4 \frac{\gh^2}{\gamh}n }{1 + 8 \frac{\gh^2}{\gamh^2} n}\rhn{n-1n-1}{} \right)
\end{align}
With these expressions we in hand we derive the second-order coherence function as explained in the next section.

\section{Derivation of the second-order coherence function of the two-ion phonon laser model}\label{sec:append:derivationOfG2}
In this section, we present a derivation of the second-order coherence function with the use of the expression for the heating and cooling spin levels found in the previous section.

We start with the differential equation of the diagonal element of the density matrix, reduced to the motional state.
\begin{align}
    \rhnd{nn}{}=&\rhnd{00,00;nn}{}+\rhnd{00,11;nn}{}+\rhnd{11,00;nn}{}+\rhnd{11,11;nn}{}\\
    =&-i\gh(\sqrt{n+1}\rhn{10,00;n+1n}{}-\sqrt{n+1}\rhn{01,00;nn+1}{})-i\gc(\sqrt{n}\rhn{00,10;n-1n}{}-\sqrt{n}\rhn{00,01;nn-1}{})\nonumber\\
    &-i\gh(\sqrt{n+1}\rhn{10,11;n+1n}{}-\sqrt{n+1}\rhn{01,11;nn+1}{})-i\gc(\sqrt{n+1}\rhn{00,01;n+1n}{}-\sqrt{n}\rhn{00,10;nn+1}{})\nonumber\\ 
    &-i\gh(\sqrt{n}\rhn{01,00;n-1n}{}-\sqrt{n+1}\rhn{10,00;nn-1}{})-i\gc(\sqrt{n}\rhn{11,10;n-1n}{}-\sqrt{n}\rhn{11,01;nn-1}{})\nonumber\\
    &-i\gh(\sqrt{n+1}\rhn{01,11;n-1n}{}-\sqrt{n+1}\rhn{10,11;nn-1}{})-i\gc(\sqrt{n}\rhn{11,01;n+1n}{}-\sqrt{n}\rhn{11,10;nn+1}{})\\
    =&-i\gh(\sqrt{n+1}\rhn{10;n+1n}{h}-\sqrt{n+1}\rhn{01;n+1n}{h}+\sqrt{n}\rhn{01;n-1n}{h}-\sqrt{n}\rhn{01;nn-1}{h})\nonumber\\
    &-i\gc(\sqrt{n}\rhn{10;n-1n}{c}-\sqrt{n}\rhn{01;nn-1}{c}+\sqrt{n+1}\rhn{01;n+1n}{c}-\sqrt{n+1}\rhn{10;nn+1}{c})\label{eq:append:eom:motion:nn}
\end{align}
As we have restricted ourselves to the diagonal of the density matrix we use in the following the shorter notation $p(n)$ for $\rhn{nn}{}$. We now use the expressions which we have found in the previous chapter and substitute them into Eq.~\eqref{eq:append:eom:motion:nn}. We find the following expression:
\begin{align}
    \dot{p}(n)=&\frac{4\frac{\gh^2}{\gamh} n}{1 + 8 \frac{gh^2}{\gamh^2} n}p(n-1)-\frac{4 \gc^2 n}{\gamc^2 + 8\gc^2 n} \left(\left(\gamc  - \frac{4 \frac{\gh^2}{\gamh}(n+1) }{1 + 8 \frac{\gh^2}{\gamh^2} (n+1)}\right)p(n)+\frac{4 \frac{\gh^2}{\gamh} n }{1 + 8 \frac{gh^2}{\gamh^2} n }p(n-1)\right)\nonumber\\
    &- \frac{4\frac{\gh^2}{\gamh} (n+1)}{1 + 8 \frac{gh^2}{\gamh^2} (n+1)}p(n)-\frac{4 \gc^2 (n+1)}{\gamc^2 + 8\gc^2 (n+1)} \left(\left(\gamc  - \frac{4 \gh^2(n+2) }{1 + 8 \frac{\gh^2}{\gamh^2} (n+2)}\right)p(n+1)+\frac{4 \frac{\gh^2}{\gamh} (n+1) }{1 + 8 \frac{gh^2}{\gamh^2} (n+1) }p(n)\right)
\end{align}

Note that the first and second lines of the right-hand side correspond to one another by replacing $n$ with $n+1$. As we are interested in the steady state distributions of the system, i.e. $\dot{p}(n)=0$, we are looking for distributions $p(n)$, which set one of the two lines to zero (which also sets the second line to zero). We therefore find a recurrence condition for $p(n)$ given by:
\begin{align}
    p(n)\frac{4 \gc^2 n}{\gamc^2 + 8\gc^2 n}\left(\gamc  - \frac{4 \frac{\gh^2}{\gamh}(n+1) }{1 + 8 \frac{\gh^2}{\gamh^2} (n+1)}\right)=p(n-1)\frac{4\frac{\gh^2}{\gamh} n}{1 + 8 \frac{gh^2}{\gamh^2} n}\left(1-\frac{4 \gc^2 n}{\gamc^2 + 8\gc^2 n}\right)
\end{align}

From this we find an expression for $p(n)$:
\begin{align}
    p(n)=p(0)\prod_{k=1}^{n}\frac{p(k)}{p(k-1)}=p(0)\prod_{k=1}^{n}\frac{\frac{\gh^2}{\gamh}(1+8\frac{\gh^2}{\gamh^2}(k+1))(1+4\frac{\gc^2}{\gamc^2}k)}{\frac{\gc^2}{\gamc}(1+8\frac{\gh^2}{\gamh^2}k)(1+8\frac{\gh^2}{\gamh^2}(k+1)-4\frac{\gh^2}{\gamh\gamc}(k+1))}
\end{align}

This allows us to calculate the two expectation values $\expval{n}$ and $\expval{n^2}$ with the following identities:
\begin{align}
    \expval{n}&=p(0)\sum_\mathrm{n=0}^\mathrm{\infty}p(n)n,\\
    \expval{n^2}&=p(0)\sum_\mathrm{n=0}^\mathrm{\infty}p(n)n^2.
\end{align}
Note that $p(0)$ is given by the normalization condition of $p(n)$.

We now calculate the second-order coherence function:

\begin{align}
    \gzn=&\frac{\expval{n^2}-\expval{n}}{\expval{n}^2}\\
    =&\left(2\left[\frac{\gc^2}{\gamc^2} (1 + 8 \frac{\gh^2}{\gamh^2})(1+ 8(2 -\frac{\gamh}{\gamc})\frac{\gh^2}{\gamh^2}) f_1 + 8\frac{\gamh}{\gamc}(1 + 4 \frac{\gc^2}{\gamc^2})\frac{\gh^4}{\gamh^4} f_2\right]\left[\left(1 - (\frac{\gh^2}{\gamh^2}\frac{\gamc^2}{\gc^2} + 8 \frac{\gh^2}{\gamh^2}) \right) f_3-  f_2\right]\right)\nonumber\\
    &\times\left((\frac{\gamh}{\gamc}-1)(1 + 4 \frac{\gc^2}{\gamc^2}) \left(\frac{\gh}{\gamh} (1 + 8 \frac{\gh^2}{\gamh^2}) f_2 + 8 \frac{\gh^3}{\gamh^3} f_4\right)^2\right)^{-1},\label{eq:append:gzn:full}
\end{align}

where we have shortened the notation for the hypergeometric functions $f_{1-4}$:

\begin{align}
    f_1&=\ _2F_1(1, 1 + 
    \frac{\gamc^2}{4\gc^2}; 2 + \frac{\gamc \gamh^2}{8\gamc\gh^2-4\gamh\gh^2}, \frac{\gamh}{2\gamc -\gamh}),\\
    f_2&=\ _2F_1(2, 2 + \frac{\gamc^2}{4 \gc^2}; 3 + \frac{\gamc \gamh^2}{8 \gamc \gh^2 - 4 \gamh \gh^2}, \frac{\gamh}{2 \gamc - \gamh}),\\
    f_3&=\ _2F_1(3, 2 + \frac{\gamc^2}{4 \gc^2}; 3 + \frac{\gamc \gamh^2}{8 \gamc \gh^2 - 4 \gamh \gh^2}, \frac{\gamh}{2 \gamc - \gamh}),\\
    f_4&=\ _pF_q(\{2, 2, 2 + \frac{\gamc^2}{4 \gc^2}\}, \{1, 3 + \frac{\gamc \gamh^2}{8 \gamc \gh^2 - 4 \gamh \gh^2}\}, \frac{\gamh}{2 \gamc - \gamh}).
\end{align}

Note that the last argument of all four functions is the same: $\frac{\gamh}{\gamc}(2-\frac{\gamh}{\gamc})^{-1}$. By using a Taylor approximation in $\gamh/\gamc$ we set the argument of the hypergeometric functions to zero to lowest order. This in turn renders all of these functions identical to one: $f_{1-4}=1$. By using the same Taylor expansion in terms of $\gamh/\gamc$ also in Eq.~\eqref{eq:append:gzn:full} we find the following expression for $\gzn$ to lowest order in $\gamh/\gamc$:

\begin{align}
    \gzn= 2-8\frac{  \frac{2\gh^2}{\gamh^2}-\frac{\gc^2}{\gamc^2}}{\left(1 +4\frac{\gc^2}{\gamc^2}\right)\left(1 + 16\frac{\gh^2}{\gamh^2}\right)}+\mathcal{O}\left(\frac{\gamh}{\gamc}\right),\ \gamh\ll\gamc
\end{align}
With this, we have found Eq.~\eqref{eq:qt:slm:g2} in the main text.\newline
We Taylor expand in $\gh\ll\gamc$, which was used in the derivation of the above result. This gives us the following expression for $\gzn$:

\begin{align}
    \gzn= \left(1 + \frac{1}{1+16\frac{\gh^2}{\gamh^2}}\right) \left(1 + 4\frac{\gc^2}{\gamc^2}\right)+\mathcal{O}\left(\frac{\gc^3}{\gamc^3}\right),\ \gc\ll\gamc
\end{align}

Lastly, for an easier analysis, we also derive an expression where $\gh\ll\gamh$. 

\begin{align}
    \gzn\approx 2   - 8\left(\frac{2\gh^2}{\gamh^2}-\frac{\gc^2}{\gamc^2}\right),\ \gc\ll\gamc,\ \gh\ll\gamh
\end{align}
This concludes the derivation of an analytic expression for $\gzn$ for the two-ion phonon lasing model.
\section{Derivation of the second-order coherence function of the single-ion model}\label{sec:append:derivationOfG2Single}
In this section, we present the derivation of the analytic expression for the second-order coherence function for the single-ion laser. We start with the special case, where $\gamh=\gamc$, discussed in subsection \ref{subsec:G2specialcase} and then in the second part broadly outline the derivation of the general case (section \ref{subsec:G2generalcase}). For convenience, we restate the Hamiltonian of the model which takes the following form in the appropriate rotating frame:
\begin{align}
    H_\mathrm{h} &= \gh(\ket{2}\bra{0} a^{\dagger} + \ket{0}\bra{2} a),\\
    H_\mathrm{c} &= \gc(\ket{1}\bra{0} a + \ket{0}\bra{1} a^{\dagger}).\\
\end{align}
The two dissipation channels are described by the following Lindbald jump operators:
\begin{align}
    L_\mathrm{h}=\sqrt{\gamh}\ket{0}\bra{2},\\
    L_\mathrm{c}=\sqrt{\gamc}\ket{0}\bra{1}.
\end{align}
Starting from this we write down the evolution equation of the ion levels evaluated at a certain Fock level $n$ and $n'$.
\begin{align}
    \rhnd{00;nn'}{}=&-i\gh(\sqrt{n+1} \rhn{10;n+1n'}{}-\rhn{01;nn'+1}{}\sqrt{n'+1}) - i \gc (\sqrt{n} \rhn{20;n-1n'}{}-\rhn{02;nn'-1}{}\sqrt{n'})\nonumber\\
    &+\gamh \rhn{11;nn'}{} +\gamc\rhn{22;nn'}{},\label{eq:singleIonG2:00}\\
    \rhnd{01;nn'+1}{}=&-i\gh(\sqrt{n+1} \rhn{11;n+1n'+1}{}-\rhn{00;nn'}{}\sqrt{n'+1})- i \gc \sqrt{n} \rhn{21;n-1n'+1}{} - \frac{\gamh}{2} \rhn{01;nn'+1}{}, \\
    \rhnd{10;n+1n'}{}=&-i\gh(\sqrt{n+1} \rhn{00;nn'}{}-\rhn{11;n+1n'+1}{}\sqrt{n'+1})+ i \gc \sqrt{n'} \rhn{12;n+1n'-1}{} - \frac{\gamh}{2} \rhn{10;n+1n'}{}, \\ 
    \rhnd{11;n+1n'+1}{}=&-i\gh(\sqrt{n+1} \rhn{01;nn'+1}{}-\rhn{10;n+1n'}{}\sqrt{n'+1}) - \gamh \rhn{11;n+1n'+1}{},\\
    \rhnd{02;nn'-1}{}=&-i\gh\sqrt{n+1} \rhn{12;n+1n'-1}{} -i \gc (\sqrt{n} \rhn{22;n-1n'-1}{}-\rhn{00;nn'}{}\sqrt{n'}) - \frac{\gamc}{2} \rhn{02;nn'-1}{}, \\ 
    \rhnd{20;n-1n'}{}=&i\gh\sqrt{n'+1} \rhn{21;n-1n'+1}{}- i \gc(\sqrt{n} \rhn{00;nn'}{}-\rhn{22;n-1n'-1}{}\sqrt{n'}) - \frac{\gamc}{2} \rhn{20;n-1n'}{},  \\
    \rhnd{22;n-1n'-1}{}=&-i\gc(\sqrt{n}\rhn{02;nn'-1}{}-\rhn{20;n-1n'}{}\sqrt{n'})  - \gamc \rhn{22;n-1n'-1}{},\\
    \rhnd{12;n+1n'-1}{}=&-i\gh\sqrt{n+1} \rhn{02;nn'-1}{} + i \gc \sqrt{n'} \rhn{10;n+1n'}{} - \frac{\gamc + \gamh}{2} \rhn{12;n+1n'-1}{},\\
    \rhnd{21;n-1n'+1}{}=&i\gh\sqrt{n'+1} \rhn{20;n-1n'}{} - i \gc \sqrt{n} \rhn{01;nn'+1}{} - \frac{\gamc + \gamh}{2} \rhn{21;n-1n'+1}{}.
\end{align}
We are faced with the problem of a non-closing set of differential equations. In order to get around this issue we would like to use the definition of the trace: $\rhn{nn'}{}=\rhn{00;nn'}{}+\rhn{11;nn'}{}+\rhn{22;nn'}{}$ and use it to substitute the last two terms in equation Eq.~\eqref{eq:singleIonG2:00}. However, this is only possible if $\gamh=\gamc$. We therefore make the assumption that $\gamh\approx\gamc$ and with this find the following approximate evolution equation for the level $\rhn{00;nn'}{}$:
\begin{align}
    \rhnd{00;nn'}{}\approx&-i\gh(\sqrt{n+1} \rhn{10;n+1n'}{}-\rhn{01;nn'+1}{}\sqrt{n'+1}) - i \gc (\sqrt{n} \rhn{20;n-1n'}{}-\rhn{02;nn'-1}{}\sqrt{n'})\nonumber\\
    &-\frac{\gamh  +\gamc}{2}\rhn{00;nn'}{}+\frac{\gamh  +\gamc}{2}\rhn{nn'}{}.\label{eq:app:singleIon:00:approx}\\
\end{align}
Note that the above equation is exact in the case where $\gamc=\gamh$. We will in the following first part assume that this is the case and rename $\gamma:=\gamh=\gamc$. With this and the dynamical equations for the other levels, we have a closing set of differential equations, which can be solved under the assumption, that $\rhn{nn'}{}$ is constant in time. The method used for this is the same as in the two-ion case. We now write down the evolution equation for a general Fock level $n$, $n'$ with the Lindblad master equation and find:
\begin{align}
    \rhnd{nn'}{} =&-i\gh (\sqrt{n}\rhn{01;n-1n'}{} - \sqrt{n'} \rhn{10;nn'-1}{}+ \sqrt{n + 1}\rhn{10;n+1n'}{} - \sqrt{n' + 1} \rhn{01;nn'+1}{} )\nonumber\\
    &-i\gc(\sqrt{n + 1} \rhn{02;n+1n'}{}-  \sqrt{n' + 1}\rhn{20;nn'+1}{} + \sqrt{n}  \rhn{20;n-1n'}{} - \sqrt{n'} \rhn{02;nn'-1}{} )\label{eq:app:singleIon:nn'}.
\end{align}
By using the found expressions for the ion levels from our previous calculation we find the following differential equation for a diagonal element of the reduced density operator ($p(n)=\rhn{nn}{}$):
\begin{align}
    \dot{p}(n)=\frac{4\frac{\gh^2}{\gamma}np(n-1)}{1+8\frac{\gc^2}{\gamma^2}(n-1)+8\frac{\gh^2}{\gamma^2}n}-\frac{4(\frac{\gh^2}{\gamma}(n+1)+\frac{\gc^2}{\gamma}n)p(n)}{1+8\frac{\gc^2}{\gamma^2}n+8\frac{\gh^2}{\gamma^2}(n+1)}+\frac{4\frac{\gc^2}{\gamma}(n+1)p(n+1)}{1+8\frac{\gc^2}{\gamma^2}(n+1)+8\frac{\gh^2}{\gamma^2}(n+2)}.
\end{align}
By invoking the steady state condition we split the resulting recurrence equation into two equivalent equations, once for $n$ and once for $n+1$. The solution to these equations is given by:
\begin{align}
    p(n)&=p(0)\prod_{k=0}^{n-1}\frac{\gh^2}{\gc^2}\frac{1+8\frac{\gc^2}{\gamma^2}(k+1)+8\frac{\gh^2}{\gamma^2}(k+2)}{1+8\frac{\gc^2}{\gamma^2}k+8\frac{\gh^2}{\gamma^2}(k+1)}=p(0)\frac{\left(\frac{\gh^2}{\gc^2}\right)^n (1+8\frac{\gc^2}{\gamma^2}k+8\frac{\gh^2}{\gamma^2}(k+1))}{1+8\gh^2}.
\end{align}
The occupation in the lowest state $p(0)$ is found through the normalization condition:
\begin{align}
    p(0)&=\left(\sum_{n=0}^{\infty}\frac{\left(\frac{\gh^2}{\gc^2}\right)^n (1+8\frac{\gc^2}{\gamma^2}k+8\frac{\gh^2}{\gamma^2}(k+1))}{1+8\gh^2}\right)^{-1}=\frac{ \frac{\gc^4}{\gamma^4} - \frac{\gc^2 \gh^2}{\gamma^4} + 16 \frac{\gc^4 \gh^2}{\gamma^6}}{(\frac{\gc^2}{\gamma^2} - \frac{gh^2}{\gamma^2})^2 (1 + 8 \frac{\gh^2}{\gamma^2})}.
\end{align}
With this we now calculate moments such as $\expval{a^\dagger a}$ and the second-order coherence function $\gzn$.
\begin{align}
    \expval{a^\dagger a}&=\frac{\gh^2 (\gamma^2 (\gc^2 - \gh^2) + 8 (\gc^4 + 3 \gc^2 \gh^2))}{(\gc^2-\gh^2) (16 \gc^2 \gh^2 + \gamma^2 (\gc^2 - \gh^2))}
\end{align}
And finally the second-order coherence function:
\begin{align}
    \gzn=\frac{\expval{n^2}-\expval{n}}{\expval{n}^2}=2 \frac{(\gamma^2 (\gc^2 - \gh^2)+ 16 \gc^2 \gh^2)(\gamma^2 (\gc^2 - \gh^2) + 16 (\gc^4 + 2 \gc^2 \gh^2))}{(\gamma^2 (\gc^2 - \gh^2)+8 (\gc^4 + 3 \gc^2 \gh^2))^2}.
\end{align}
With this, we have found the expression Eq.~\eqref{eq:singleIon:g2} in the main text.

We now consider the more general case, where $\gamh\approx\gamc$. We can still write down the equation Eq.~\eqref{eq:app:singleIon:00:approx}, which in this case holds only approximately. The coupled differential equation can still be solved and substituted into Eq.~\eqref{eq:app:singleIon:nn'}. This will result in two equivalent difference equations which can again be solved as in the case where $\gamh=\gamc$. With this, we find an expression for $p(n)$ as well as for $\gzn$. However, now the resulting expressions are considerably more complex and will not be given here in detail. We would expect the results to hold in case where $\gamh\approx\gamc$. As a comparison with the simulation shows, this holds even when this condition is not fulfilled anymore.
\section{Derivation of higher-order Lamb-Dicke terms}
In this section, we derive the higher-order terms which appear if we leave the regime, where the first-order Lamb-Dicke approximation is valid. We start this derivation with the interaction Hamiltonian of the laser field and the ion in the harmonic trap
\begin{align}
    H_\mathrm{I}=\Omega\left(\sigma_\mathrm{+}+\sigma_\mathrm{-}\right)\left(e^{i(kz-\omega t +\phi)}+e^{-i(kz-\omega t +\phi)}\right),
\end{align}
where $\omega$ and $\phi$ are the frequency and phase of the laser drive, $\Omega$ is the Rabi frequency and determined by the geometry of the setup and the laser drive. By moving into the interaction picture with respect to the ion trap and ion spin Hamiltonian, we obtain the following interaction Hamiltonian
\begin{align}
    H_\mathrm{I}=\Omega\left(\sigma_\mathrm{+}e^{i\omega_\mathrm{q}t}+\sigma_\mathrm{-}e^{-i\omega_\mathrm{q}t}\right)\left(e^{i(\eta(a e^{i\omega_\mathrm{m} t} + a^\dagger e^{-i\omega_\mathrm{m} t})-\omega t +\phi)}+e^{-i(\eta(a e^{i\omega_\mathrm{m} t} + a^\dagger e^{-i\omega_\mathrm{m} t})-\omega t +\phi)}\right),
\end{align}
where we have written $kz= \abs{k}z_\mathrm{0}(a + a^\dagger)$ and $\omega_\mathrm{m/q}$ is the motional frequency of the trap and the qubit frequency respectively. With a rotating wave approximation, we end up with
\begin{align}
    H_\mathrm{I}=\Omega\left(\sigma_\mathrm{+}e^{i(\eta(a e^{i\omega_\mathrm{m} t} + a^\dagger e^{-i\omega_\mathrm{m} t})-\delta t +\phi)}+\sigma_\mathrm{-}e^{-i(\eta(a e^{i\omega_\mathrm{m} t} + a^\dagger e^{-i\omega_\mathrm{m} t})-\delta t +\phi)}\right),
\end{align}
where $\delta = \omega_\mathrm{q}-\omega_\mathrm{m}$. We now use the following identity to simplify the above expression:
\begin{align}
    e^{i(\eta(a e^{i\omega_\mathrm{m} t} + a^\dagger e^{-i\omega_\mathrm{m} t}))}=e^{\frac{-\eta^2}{2}}\sum_\mathrm{n,m}^{\infty}\frac{(i\eta a^\dagger e^{i\omega_\mathrm{m}t})^n(i\eta a e^{i\omega_\mathrm{m}t})^m}{n!m!}.
\end{align}
With this and by setting $\delta = \pm \omega_\mathrm{m}$ we get an expression for the two sideband Hamiltonians:
\begin{align}
    H_\mathrm{BSB}&=\Omega_\mathrm{h} e^{\frac{-\eta^2}{2}}\eta\sum_\mathrm{j=0}^{\infty}\frac{(-\eta^2)^j}{(j+1)!j!}\left(\sigma_\mathrm{+}(a^\dagger)^{j+1}(a)^j+\sigma_\mathrm{-}(a^\dagger)^{j}(a)^{j+1}\right),\\
    H_\mathrm{RSB}&=\Omega_\mathrm{c} e^{\frac{-\eta^2}{2}}\eta\sum_\mathrm{j=0}^{\infty}\frac{(-\eta^2)^j}{(j+1)!j!}\left(\sigma_\mathrm{+}(a^\dagger)^{j}(a)^{j+1}+\sigma_\mathrm{-}(a^\dagger)^{j+1}(a)^{j}\right).
\end{align}
We identify $\gh=\Omega_\mathrm{h} e^{\frac{-\eta^2}{2}}\eta$ and $\gc=\Omega_\mathrm{c}e^{\frac{-\eta^2}{2}}\eta$. From this, we recover the usual blue and red sideband Hamiltonians for $j=0$ and then the higher-order terms for increasing powers of $\eta$.

\section{Derivation of Fisher information for a squeezed phonon laser}\label{sec:append:derivationOfFisher}
We present a derivation of the Fisher information for the case of lasing in a squeezed mode, following the approach outlined in Ref.~\cite{Porras2017}. We start with the lasing Hamiltonian of the heating ion and the squeezed bosonic mode $A$, the adiabatically eliminated cooling ion and the applied signal Hamiltonian
\begin{align}
    H_\mathrm{h,sq} &= g_\mathrm{h}(A^\dagger\sph+A\smh),\\
    L_{\gamma} &=\sqrt{\gamh}\smh,\\
    L_{\kappa,sq} &=\sqrt{\kappa_\mathrm{c}}A,\\
    H_\mathrm{signal} &= \epsilon^* a + \epsilon a^\dagger,
\end{align}
where $A=S(\xi)aS^\dagger(\xi)$ and $S(\xi)=e^{(\xi a^2 - \xi (a^\dagger)^2)/2}$. We find the equations of motion using the master equation in Lindblad from and trace out the spin degree of freedom. By expanding in orders of $\gh/\gamh$ and neglecting third and higher order terms, we end up with the modified evolution equation for the reduced density matrix:
\begin{align}
    \dot{\rho_f}=&-(\epsilon^*a\rho_f+\epsilon a^\dagger\rho_f-\epsilon^*\rho_fa-\epsilon\rho_fa^\dagger) \nonumber\\
    &+\frac{16\gh^4}{\gamh^3}(AA^\dagger \rho_f A A^\dagger - (A^\dagger)^2 \rho_f A^2)\nonumber \\
    &+ \frac{2\gh^2}{\gamh}( 2A^\dagger\rho_f A -AA^\dagger\rho_f-\rho_fAA^\dagger)\nonumber \\
    &+\kapc A\rho_fA^\dagger - \frac{\kapc}{2}(A^\dagger A \rho_f +\rho_fA^\dagger A).\label{eq:derivation:rhof:final}
\end{align}
Analogous to the Fokker-Planck method, we translate this equation into a partial differential equation, by expressing the density matrix in terms of a quasi probability distribution P 
\begin{align}
    \rho=\int d^2\alpha P(\alpha,\alpha^*,t) S(\xi)\ket{\alpha}\bra{\alpha}S^\dagger(\xi).
\end{align}
Following the same derivation as outlined in Ref.~\cite{scully_zubairy_1997,Porras2017}, we get the following expression for the steady-state
\begin{align}
    P(I,\theta)=\frac{1}{N}\exp\left[-\frac{B}{2A}I^4+\frac{A-C}{A}I^2-\frac{2\abs{\delta}}{A}I\sin(\theta -\gamma)\right],
\end{align}
where $A=\frac{2\gh^2}{\gamh}$, $B=\frac{16\gh^4}{\gamh^3}$, $C=\kapc$ and
\begin{align}
    \delta:= \cosh(r)\epsilon - e^{i\beta}\sinh(r)\epsilon^*=\abs{\delta}e^{i\gamma}.
\end{align}
Note that the squeezing parameter is given by ($\xi=re^{i\beta}$), as in the main text. The optimal measurement operator for determining the amplitude of the external signal $\epsilon$ is given by
\begin{align}
    P_\gamma = -i(Ae^{-i\gamma}-A^\dagger e^{i\gamma}).
\end{align}
The Fisher information of such a measurement is given by
\begin{align}
    F_Q[\rho_{\abs{\epsilon}}]\approx\frac{2I^2}{A^2}W^2,
\end{align}
where $W=\cosh^2(r)+\sinh^2(r)-\cosh(r)\sinh(r)\cos(\beta-2\phi)$. In the optimal case, the squeezing and the phase of the applied signal are orthogonal to one another. In this case we improve the sensitivity with $\propto 1/(\cosh^2(r)+\sinh^2(r))^2$.

\section{Additional simulation results for single-ion lasing scheme}\label{sec:append:singleIonSim}
In this section we provide additional simulation results for the single-ion phonon laser.
\subsection{Higher-order Lamb-Dicke approximation}
\begin{figure}[h!]
    \centering
    \includegraphics[width=0.4\textwidth]{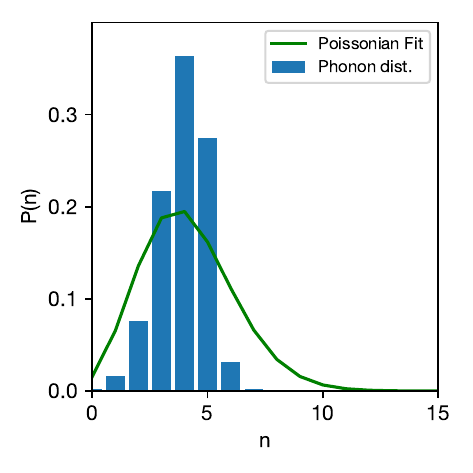}
    \caption{The steady-state phonon distribution (blue bars) simulated with up to third-order Lamb-Dicke terms. The green line indicates a Poissonian distribution, signifying a coherent state. We find a sub-Poissonian phonon distribution and therefore a non-classical state.}
    \label{fig:SuppFigBeyondLD}
\end{figure}

We present in Fig.~\ref{fig:SuppFigBeyondLD} the simulation results for an exemplary steady state of the single-ion phonon laser in the Lamb-Dicke regime, where higher order terms of the LD expansion have to be taken into account. The steady state shows similar sub-Poissonian statistics as in the two-ion case.
\subsection{Squeezing}
\begin{figure}[h!]
    \centering
    \includegraphics[width=0.45\textwidth]{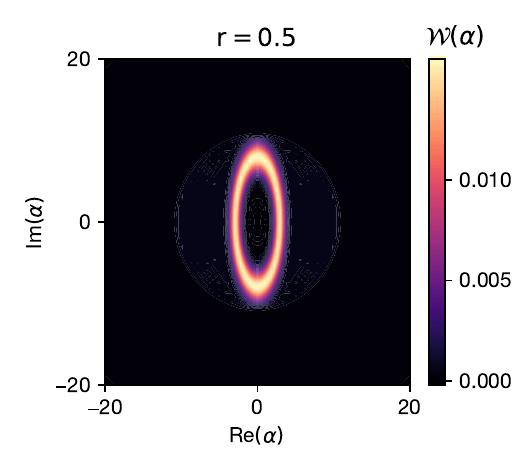}
    \caption{Simulated Wigner function $\mathcal{W}(\alpha)$ for the steady states of the phonon laser, in the phonon field phase space $\alpha$ with a squeezing parameter of $r=0.8$. We have chosen the squeezing phase to be zero ($\beta=0$). We therefore see that the real quadrature of the phonon mode gets squeezed, while the imaginary is anti-squeezed.}
    \label{fig:SuppFigsqueezing}
\end{figure}

In Fig.~\ref{fig:SuppFigsqueezing} we show an exemplary steady state of the single-ion phonon laser if the squeezing scheme as discussed in the main text is applied. We again find a qualitatively similar Wigner function, as in the two-ion lasing scheme.

\end{document}